\documentclass{article}
\usepackage{graphicx}
\usepackage{amsmath, amssymb}
\usepackage{mathtools}
\usepackage{bbm}
\usepackage[numbers]{natbib}
\usepackage[normalem]{ulem}
\usepackage{subcaption}
\usepackage{standalone}
\usepackage{booktabs}
\usepackage{longtable}
\usepackage{pdflscape}
\usepackage{placeins}
\usepackage{floatpag}
\usepackage{url}

\usepackage{algorithm}
\usepackage{algpseudocode}
\usepackage{graphicx} % Required for % Required for inserting images
\usepackage{xcolor} % Required for % Required for inserting images

\title{
Fast simulation of nonlinear deep resistive networks for energy-based computation}

\author{
\begin{tabular}{c}
{\small
Filip Osana\textsuperscript{(1)(2)}, 
Julie Grollier\textsuperscript{(2)*},
Damien Querlioz\textsuperscript{(1)}
}
\\[0.4em]
{\footnotesize
\textsuperscript{(1)} Université Paris-Saclay, CNRS, Centre de Nanosciences et de Nanotechnologies, Palaiseau, France
}
\\
{\footnotesize
\textsuperscript{(2)} Laboratoire Albert Fert, CNRS, Thales, Université Paris-Saclay, Palaiseau, France
}
\\
{\footnotesize
\textsuperscript{*} Corresponding author: julie.grollier@cnrs-thales.fr
}
\end{tabular}
}

\date{}

\begin{document}
\maketitle
\begin{abstract}

Deep resistive networks are electronic energy-based systems in which computation is performed by the steady-state voltages of nonlinear circuits. Nonlinear devices enable expressive input-output transformations, but make the circuit equilibria costly to compute during simulation and training. Recent coordinate-descent solvers have achieved large speedups over SPICE-class circuit simulators, but only for nonlinearities modeled as ideal-diode models. This mathematical simplification is not sufficient for practical analog circuits. Here we extend coordinate-descent simulation to realistic monotone nonlinearities, including Shockley diodes, antiparallel diode pairs, and piecewise-linear current-voltage characteristics. With neighboring voltages fixed, each node update remains a scalar Kirchhoff-law solve. Single-exponential characteristics admit closed-form Lambert-(W) updates, while more general monotone characteristics can be handled with  scalar root-finding methods. Across networks with one to three hidden layers and hidden widths from 64 to 1024, the solver reproduces matched SPICE steady-state voltages with relative $L_1$ errors below $1.1\times10^{-4}$ for 90\% of validation samples, while achieving speedups up to $(1.7\times10^3)$. We further train a $1568\times100\times20$ double-Shockley network on MNIST, reaching about 3.0\% test error and reducing the per-epoch training time by roughly $4.4\times10^2$. Together, these results establish a practical route to the circuit-level design and training of large-scale analog energy-based systems incorporating realistic nonlinear devices.

\end{abstract}

\section*{Introduction}

\renewcommand\labelitemi{$\bullet$}
\renewcommand\labelitemii{$\circ$}
\renewcommand\labelitemiii{$\triangleright$}
\renewcommand\labelitemiv{$\cdot$}
\label{sec:introduction}

Physical machines for energy-based computation use the dynamics of physics to perform part of the computation: a state evolves under physical laws until it reaches an equilibrium, attractor, or minimum of an energy-like landscape \citep{Markovic2020,Xiao2020,XiaYang2019,christensen20222022,momeni2025training}. In electronic implementations, this idea is especially appealing because Kirchhoff's laws, device nonlinearities, and dissipation are then mechanisms that perform energy minimization intrinsically. %The challenge is that such machines must still be designed, trained, and benchmarked before they can be built at useful scale.

Deep resistive networks (DRNs), whose topology is inspired by deep neural networks, provide a promising electronic implementation of energy-based computation \cite{Kendall2020,ScellierMishra2023}. In these circuits, trainable parameters are encoded by conductances connecting successive layers, while nonlinear one-port elements play the role of activation functions. The node voltages are determined by Kirchhoff's current law, and for monotone nonlinear elements the steady state can be written as the minimizer of an electrical content function, which plays the role of the ``energy function'' of the network. This energy-minimization formulation makes DRNs compatible with equilibrium propagation \cite{ScellierBengio2017,scellier2021deep,laborieux2021scaling}, in which learning is obtained from local differences between free and nudged equilibria. In principle, this gives a route to trainable analog physical learning machines \citep{Dillavou2022,Dillavou2023}.

A central challenge, however, is that these networks must be designed, trained, and benchmarked in simulation before they can be built at useful scale. A single inference requires solving for the steady-state voltages of the circuit, each training step requires one or more steady-state solves, and a complete training run can require millions of such solves. This makes DRNs particularly hard to study and optimize. General-purpose circuit simulators such as SPICE provide reliable reference solutions for DRNs, but they are slow for the repeated steady-state solves required in inference and learning. Each solve is treated as a generic nonlinear circuit simulation, without exploiting the layered structure of DRNs or the fact that many similar equilibria must be computed. This computational cost has limited SPICE-based studies of nonlinear DRNs to modest network sizes \citep{Kendall2020,Watfa2023}. More broadly, scalable hardware demonstrations of resistive learning systems have often avoided this nonlinear-equilibrium bottleneck by focusing on linear networks \citep{Stern2022,Stern2024,Wycoff2022}.
%This has limited SPICE-based studies to modest network sizes \citep{Kendall2020,Watfa2023}, while many scalable demonstrations have relied on linear networks \citep{Stern2022,Stern2024,Wycoff2022}.

Recent work showed that coordinate descent can greatly accelerate DRN simulation when the nonlinear elements are modeled as ideal diodes \citep{Scellier2024}. In that special case, the equilibrium problem reduces to a quadratic program with linear inequality constraints, which makes the coordinate updates particularly simple. However, ideal diodes are a severe simplification of physical circuit behavior. In practical analog circuits, nonlinear behavior is governed by realistic current-voltage characteristics, such as Shockley diode curves, antiparallel diode pairs, or transistor-derived responses. The challenge is therefore to retain the speed of coordinate descent while moving beyond ideal constraints to physical nonlinear device characteristics.

Realistic current-voltage characteristics can be specified either by analytical expressions or by measured tabulated data. They no longer give the quadratic program obtained for ideal-diode networks, but when the characteristics are monotone their contribution to the electrical content remains convex. The key observation underlying this work is that this is sufficient to preserve the coordinate-descent approach. With neighboring voltages held fixed, updating a node voltage still only requires enforcing Kirchhoff’s current law at that node, leading to a scalar nonlinear equation. Device realism therefore only changes the local update rule, while the layer-wise coordinate-descent structure of DRNs is preserved.

Here, we turn this observation into a coordinate-descent solver for nonlinear DRNs. We first formulate networks with linear conductances, ideal sources, and monotone nonlinear one-port devices as convex content minimization problems. We then derive local coordinate updates for analytical and tabulated nonlinear current-voltage characteristics, including closed-form Lambert-\(W\) updates for single-exponential branches and scalar updates for antiparallel Shockley and monotone piecewise-linear characteristics. We validate the resulting solver against matched SPICE simulations on Digits networks with one to three hidden layers, hidden widths from 64 to 1024, and three nonlinear device models. Finally, we use the solver to train a \(1568\times100\times20\) double-Shockley DRN on MNIST and verify the trained circuit against SPICE.

\section*{Results}
\subsection*{Resistive networks with nonlinear one-port elements as energy-based machines}
\label{sec:proposed_method_outline}

\begin{figure}[!htbp]
    \centering

    % --- top: main network ---
    \begin{subfigure}{0.8\linewidth}
        \centering
        \includegraphics[width=\linewidth]{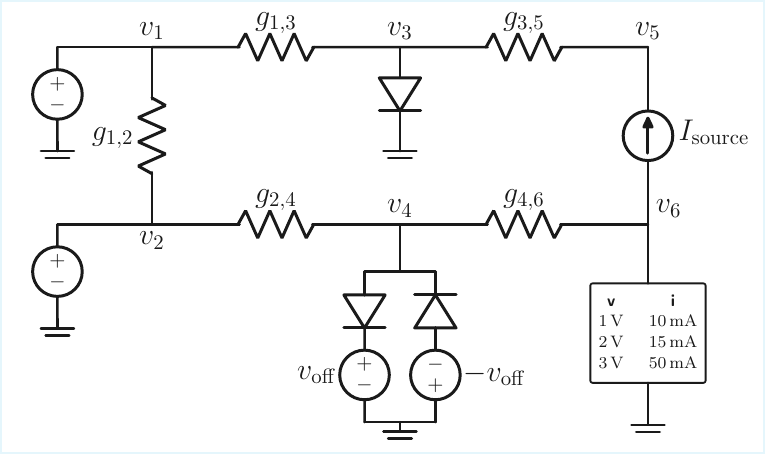}
        \caption{The general schematic of the circuit that can be solved using our CD solver.}
        \label{fig:network_schematic}
    \end{subfigure}

    \vspace{0.6em}

    % --- bottom: three nonlinearities ---
    \begin{subfigure}{0.31\linewidth}
        \centering
        \includegraphics[width=\linewidth]{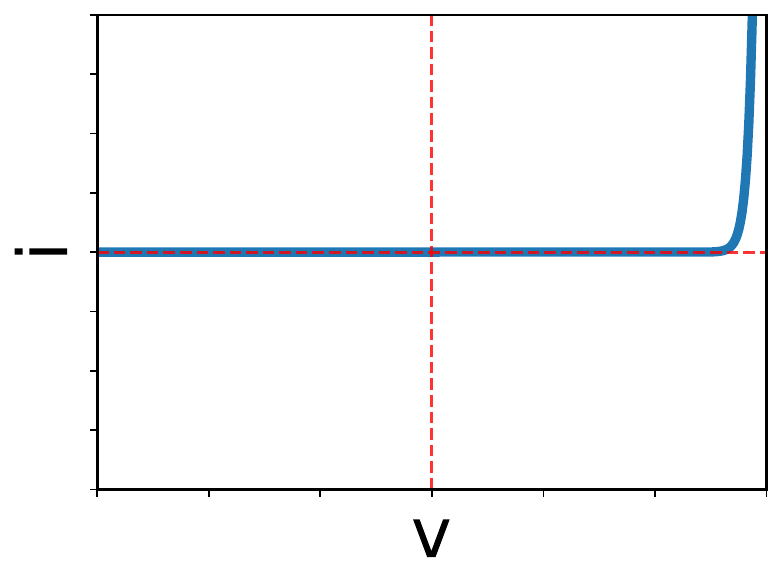}
        \caption{Single Shockley diode.}
        \label{fig:shockley}
    \end{subfigure}\hfill
    \begin{subfigure}{0.31\linewidth}
        \centering
        \includegraphics[width=\linewidth]{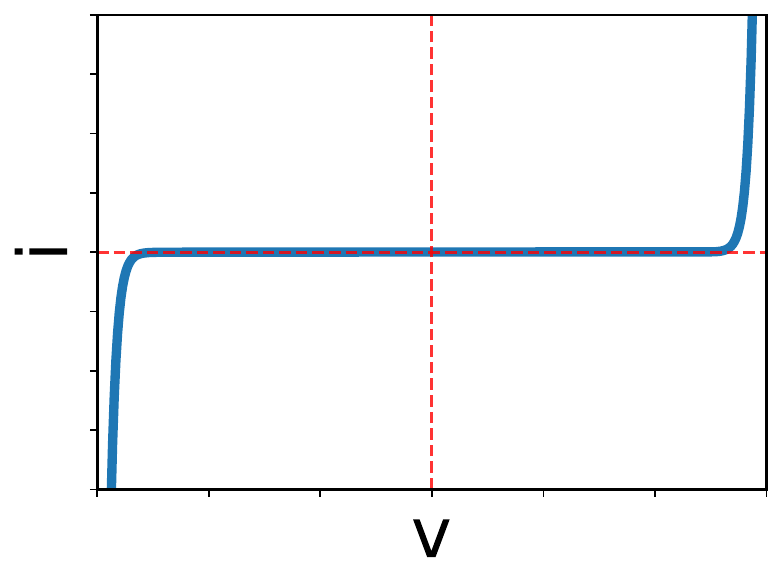}
        \caption{Double Shockley diode.}
        \label{fig:double_shockley}
    \end{subfigure}\hfill
    \begin{subfigure}{0.31\linewidth}
        \centering
        \includegraphics[width=\linewidth]{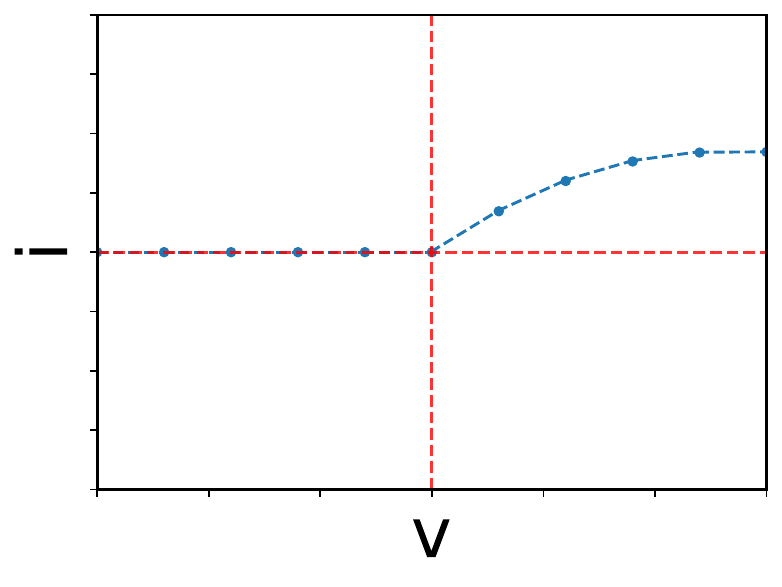}
        \caption{PWL curve.}
        \label{fig:mosfet_vds}
    \end{subfigure}

    \caption{\textbf{Network class and nonlinear device models  considered in this work.}
    (a) Representative resistive network with ideal sources, linear conductances, and nonlinear one-port elements connected to fixed potentials. In the schematic, these one-port elements are drawn with diode symbols for simplicity, but they represent generic voltage-controlled nonlinear current-voltage characteristics.
    (b-d) Device characteristics used in the experiments: Shockley diode,  double Shockley diode, and monotone piecewise-linear (PWL) $i$-$v$ curve.}
    \label{fig:network}
\end{figure}

%We consider electrical networks composed of ideal voltage sources, ideal current sources, linear resistors, and voltage-controlled nonlinear one-port elements. Each nonlinear element is attached to a free node $k$, while its other terminal is tied to a fixed potential. For notational simplicity, this fixed terminal potential is absorbed into the device law, so $i_k(v_k)$ denotes the current leaving node $k$ through the nonlinear element as a function of the corresponding node voltage.
We consider electrical networks composed of ideal voltage sources, ideal current sources, linear resistors, and voltage-controlled nonlinear one-port elements, as illustrated in Fig.~\ref{fig:network}. The linear resistors connect pairs of circuit nodes and represent the trainable conductances of the network. By contrast, the nonlinear elements are local shunt elements: they are attached to free nodes and connected on their other terminal to a fixed potential, such as ground or a voltage bias. If the fixed terminal of the nonlinear element connected to node (k) is held at $V_k^{\mathrm{fix}}$, its current depends on the voltage difference $v_k - V_k^{\mathrm{fix}}$. Since $V_k^{\mathrm{fix}}$ is prescribed, we absorb this constant offset into the definition of the device characteristic and write $i_k(v_k)$ for the current leaving node (k) through the nonlinear branch.

The network has $N$ nodes and node potentials $v=(v_1,\dots,v_N)\in\mathbb{R}^N$.
We define $\mathcal{V}\coloneqq\{1,\dots,N\}$.
We index circuit elements by the sets
\[
B_R \subseteq \bigl\{\{j,k\}\subseteq\mathcal V:\ j\neq k \bigr\}
\quad\text{(resistor branches)},
\qquad
B_{CS} \subseteq \mathcal{V}\times\mathcal{V}
\quad\text{(current sources)},
\]
\[
V_{NL} \subseteq \mathcal{V}
\quad\text{(nodes with a nonlinear element)}.
\]

Voltage sources clamp boundary nodes to prescribed potentials. We denote by $S\subseteq\mathbb{R}^N$ the set of node
potentials compatible with these constraints and a chosen reference node. We denote by
$V_{\mathrm{free}}\subseteq\mathcal{V}$ the set of free, non-boundary nodes, i.e.,  nodes whose voltages are not fixed by voltage-source constraints.

For each resistor branch $\{j,k\}\in B_R$, $g_{jk}\ge 0$ denotes the conductance between nodes $j$ and $k$.
For each current-source branch $(j,k)\in B_{CS}$, $I^{CS}_{jk}$ denotes the current leaving node $j$ and entering node $k$.
For the nonlinear branches, we restrict attention to monotone current-voltage characteristics. This assumption covers the passive nonlinear elements considered here and ensures that each branch contributes a convex term to the electrical content. Specifically, for each $k\in V_{NL}$, the nonlinear element is described by a nondecreasing current law
$i_k(\cdot)$ and its primitive function \cite{Millar1951Resistance,Cherry1951Reactance,Johnson2010NonlinearElectricalNetworks}.
\begin{equation}
\phi_k(v)\;\coloneqq\;\int_{0}^{v} i_k(u)\,\mathrm{d}u .
\label{eq:phi_node_def}
\end{equation}
Because \(i_k\) is nondecreasing, \(\phi_k\) is convex.

We define the energy function of the network as
\begin{equation}
E(v)
\;=\;
E_{\mathrm{lin}}(v)\;+\;E_{\mathrm{nl}}(v)\;+\;E_{\mathrm{cs}}(v),
\label{eq:E_split_simple}
\end{equation}
with
\begin{align}
E_{\mathrm{lin}}(v)
&\;\coloneqq\;
\frac{1}{2}\sum_{\{j,k\}\in B_R} g_{jk}\,\bigl(v_j-v_k\bigr)^2,
\label{eq:E_lin_simple}\\
E_{\mathrm{nl}}(v)
&\;\coloneqq\;
\sum_{k\in V_{NL}} \phi_k(v_k),
\label{eq:E_nl_simple}\\
E_{\mathrm{cs}}(v)
&\;\coloneqq\;
\sum_{(j,k)\in B_{CS}} I^{CS}_{jk}\,\bigl(v_j-v_k\bigr).
\label{eq:E_cs_simple}
\end{align}
The energy function of the network is called this way as it plays the role of energy in energy-based networks, although it is homogeneous to a power. In the literature, this quantity is also called content function \cite{Millar1951Resistance,Cherry1951Reactance} or pseudopower \cite{Johnson2010NonlinearElectricalNetworks}.

For a free node $k$, differentiating $E$ with respect to $v_k$ gives the residual current leaving that node:
\begin{equation}
\frac{\partial E}{\partial v_k}(v)
=
\sum_{j:\{j,k\}\in B_R} g_{jk}(v_k-v_j)
+
\begin{cases}
i_k(v_k), & k\in V_{NL},\\
0, & k\notin V_{NL},
\end{cases}
+
\sum_{j:(k,j)\in B_{CS}} I^{CS}_{kj}
-
\sum_{j:(j,k)\in B_{CS}} I^{CS}_{jk}.
\label{eq:kcl_residual_energy_gradient}
\end{equation}

Thus, the stationarity conditions $\partial E/\partial v_k=0$ on free nodes are precisely the KCL equations at free nodes.

Since the resistor contributions are convex for $g_{jk}\ge 0$, the nonlinear contributions are convex because $i_k(\cdot)$ is nondecreasing, and the current-source contributions are linear, the content $E$ is convex on $S$.
The KCL solution can be characterized as the minimizer of this electrical content 
%the KCL solution is equivalently characterized as the minimizer
\begin{equation}
v^\star \;=\;\arg\min_{v\in S} E(v),
\label{eq:variational_characterization}
\end{equation}
\cite{Millar1951Resistance,Cherry1951Reactance,Johnson2010NonlinearElectricalNetworks}.

When \(E\) is strictly convex on the free variables, this equilibrium is unique.

\subsection*{Coordinate descent reduces each update to a local KCL solve}
\label{subsec:our_cd_step_outline}

%Since $g_{jk}\ge 0$ and each $\phi_k$ is convex (because $i_k(\cdot)$ is monotonically increasing), the objective $E(v)$ is convex on the feasible set $S$.
The convexity of the content objective $E$ on $S$ allows us to solve~\eqref{eq:variational_characterization} using coordinate descent,
following the approach of ref.~\cite{Scellier2024}.
In ref.~\cite{Scellier2024}, the nonlinear elements are ideal diodes, which do not contribute in terms of the
dissipation potential and are enforced by linear inequality constraints. This assumption yields a
quadratic objective with linear inequality constraints.
Here, we generalize to the nonlinearities that contribute through convex dissipation potentials
$\phi_k$, so the objective remains convex but it is non-quadratic.

%\subsubsection{Coordinate descent}
Coordinate descent algorithms minimize the energy function of the network by iteratively updating one coordinate at a time.
At iteration $t$, we select a free node $k_t\in V_{\mathrm{free}}$.
All other node voltages $\{v_j\}_{j\neq k_t}$ are held fixed,  and \(v_{k_t}\) is updated by solving
\begin{equation}
v_{k_t}^{t+1}\in\arg\min_{u\in\mathbb{R}}\;
E\!\bigl(v_1^t,\dots,v_{k_t-1}^t,u,v_{k_t+1}^t,\dots,v_N^t\bigr).
\label{eq:exact_scalar_cd_scalar}
\end{equation}
With all other node voltages held fixed, the only terms in the energy function of the network that depend on
 the selected voltage  are  the resistor branches, current sources, and nonlinear
elements incident on node $k$. Therefore, the one-dimensional subproblem can be written as
\begin{equation}
\begin{aligned}
E_k(v_k)
\;&=\;
a_k\,v_k^2 \;+\; b_k\,v_k \;+\; \phi_k(v_k), \\
a_k \;:&=\; \frac{1}{2}\sum_{j\in\mathcal{N}_R(k)} g_{kj}, \\
b_k \;:&=\; -\sum_{j\in\mathcal{N}_R(k)} g_{kj}\,v_j
\;+\;\sum_{j:(k,j)\in B_{CS}} I^{CS}_{kj}
\;-\;\sum_{j:(j,k)\in B_{CS}} I^{CS}_{jk}.
\end{aligned}
\label{eq:Ek_general_abphi}
\end{equation}
Here, $\phi_k\equiv 0$ if node $k$ has  no nonlinear one-port element, and 
\begin{equation}
\mathcal{N}_R(k)
\;:=\;
\left\{\, j \in \mathcal{V}\setminus\{k\} \;:\; \{j,k\}\in B_R \,\right\}
\end{equation}
is the set of resistive neighbors of node $k$.

The exact coordinate update is given by the stationarity condition
$\frac{\partial E_k}{\partial v_k}(v_k)=0$, i.e.
\begin{equation}
2a_k\,v_k \;+\; b_k \;+\; \frac{\partial \phi_k}{\partial v}(v_k) \;=\; 0.
\label{eq:stationarity_Ek_general}
\end{equation}
For differentiable device laws, 
$\frac{\partial \phi_k}{\partial v}(v)=i_k(v)$, 
so the update is obtained by solving
\begin{equation}
2a_k\,v_k \;+\; b_k \;+\; i_k(v_k) \;=\; 0,
\label{eq:1d_kcl_update_general}
\end{equation}
which is the KCL equation at node $k$ with all neighboring voltages held fixed.

Therefore, an exact coordinate-descent update is equivalent to solving the KCL equation at a single node.
%During intermediate CD iterations, the free-node KCL residuals do not generally vanish simultaneously. The selected node residual is set to zero by the update, but residuals at other free nodes usually remain nonzero because their neighboring voltages have changed. At CD equilibrium, all free-node KCL equations are satisfied:
During intermediate CD iterations, only the selected node is brought into local KCL balance; the residuals at other free nodes generally remain nonzero because they depend on neighboring voltages that may have changed. At convergence, all free-node KCL equations are satisfied:
\begin{equation}
\frac{\partial E}{\partial v_k}(v^\star)=0,
\qquad \forall k\in V_{\mathrm{free}}.
\label{eq:kcl_all_nodes_cd}
\end{equation}

The efficiency of the solver therefore depends on how quickly and accurately the scalar subproblem in Eq.~\eqref{eq:1d_kcl_update_general} can be solved for each nonlinear device law. Below, we derive these scalar updates for the nonlinearities used in this work.

%To keep CD efficient, we must be able to solve each subproblem quickly. In the following we explore how to solve such subproblem for different non-linearities. 

\subsection*{Device-specific scalar updates}
\paragraph*{Shockley diode.}

We first consider the Shockley diode with current-voltage relation
  \begin{equation}
  i(v)
  \;=\;
  I_s\!\left(\exp\!\Big(\frac{v - v_{\mathrm{off}}}{v_T}\Big)-1\right).
  \qquad I_s\ge 0,\; v_T>0
  \label{eq:shockley_iv}
  \end{equation}

For this device law, the scalar KCL equation in Eq.~\eqref{eq:1d_kcl_update_general} becomes  
\begin{equation}
  \frac{\partial E}{\partial v_k} = 2a_k v_k + b_k + I_s\!\left(\exp\!\Big(\frac{v_k - v_{\mathrm{off}}}{v_T}\Big)-1\right)=0.
  \label{eq:shockley_stationarity}
  \end{equation}
This equation has the Lambert-\(W\) form
  \begin{equation}
  A\,x+B+C\,e^{D x}=0,
  \qquad A\neq 0,\; D\neq 0,
  \label{eq:shockley_lambert_generic_form}
  \end{equation}
  whose  principal branch solution is
  \begin{equation}
  x^\star
  =
  -\frac{B}{A}
  -\frac{1}{D}\,
  W_0\!\left(
  \frac{D C}{A}\,e^{-D B/A}
  \right),
  \label{eq:lambert_generic_solution}
  \end{equation}
where $W_0$ denotes the principal branch of the Lambert-\(W\) function.

Therefore, the coordinate update is available in closed form:
\begin{equation}
v_k^\star
=
-\frac{b_k-I_s}{2a_k}
-
v_T\,W_0\!\left(
\frac{I_s}{2a_k v_T}\,
\exp\!\Big(
-\frac{v_{\mathrm{off}}}{v_T}
-\frac{b_k-I_s}{2a_k v_T}
\Big)
\right).
\label{eq:supp_shockley_vstar}
\end{equation}

For $a_k>0$, $I_s>0$, and $v_T>0$,
the argument of $W_0(\cdot)$ is strictly positive,
so \eqref{eq:supp_shockley_vstar} gives a real update. As $E(v_k)$ is strictly convex,  $v_k^\star$ is unique.
The full derivation is provided in Supplementary Sec.~\ref{sec:supp_shockley_lambert}

\paragraph*{Double Shockley diode.}

We next consider the antiparallel diode nonlinearity used in ref.~\cite{Kendall2020}. The full antiparallel Shockley current-voltage law is
  \begin{equation}
  i(v)
  \;=\;
  I_s\!\left(\exp\!\Big(\frac{v - v_{\mathrm{off},1}}{v_T}\Big)-1\right)
  -
  I_s\!\left(\exp\!\Big(\frac{-(v+v_{\mathrm{off},2})}{v_T}\Big)-1\right),
  \qquad I_s\ge 0,\; v_T>0.
  \label{eq:double_diode}
  \end{equation}
where \(v_{\mathrm{off},1}>0\) and \(v_{\mathrm{off},2}>0\) are the positive and negative branch voltage offsets. 

Substituting Eq.~\eqref{eq:double_diode} into the scalar KCL (Eq.~\eqref{eq:1d_kcl_update_general}) gives
  \begin{equation}
  2a_k v_k+b_k
  \;+\;
  I_s\exp\!\Big(\frac{v_k - v_{\mathrm{off},1}}{v_T}\Big)
  \;-\;
  I_s\exp\!\Big(\frac{-(v_k+v_{\mathrm{off},2})}{v_T}\Big)
  \;=\;0,
  \label{eq:double_diode_stationarity_expanded}
  \end{equation}

The constant $-I_s$ terms cancel in the full antiparallel model. 
Eq.~\eqref{eq:double_diode_stationarity_expanded} contains both \(\exp(v_k/v_T)\) and
\(\exp(-v_k/v_T)\), so it does not reduce to the Lambert-\(W\) form  used above. 

%A numerical root-finding method, such as Newton--Raphson, could be used to solve \eqref{eq:double_diode_stationarity_expanded}. However, this would lead to an iterative solution rather than the closed-form Lambert-\(W\) update considered here.
To recover a Lambert-$W$ form
for the CD solver, we use an active-branch approximation in which only the forward-biased diode branch is retained:
%, we can introduce an approximation in which only the forward-biased diode is assumed to conduct at a time:
\begin{equation}
  i_{LS}{\mathrm{}}(v)
  \;\approx\;
  \begin{cases}
  I_s\,\exp\!\Big(\dfrac{v - v_{\mathrm{off},1}}{v_T}\Big), & v>0,\\[10pt]
  -\,I_s\,\exp\!\Big(\dfrac{-(v+v_{\mathrm{off},2})}{v_T}\Big), & v<0.
  \end{cases}
  \label{eq:double_diode_large_signal}
  \end{equation}

Away from the switching point \(v_k=0\), the coordinate update is therefore obtained from
  \begin{equation}
  \frac{\partial E}{\partial v_k}(v_k)=2a_k v_k+b_k+i_{\mathrm{LS}}(v_k)=0.
  \label{eq:double_diode_stationarity_LS}
  \end{equation}
Under Eq.~\eqref{eq:double_diode_large_signal}, the scalar objective is strictly convex, with a strictly increasing derivative on each open branch and an upward jump at \(v_k=0\). The signs of the one-sided derivatives therefore determine whether the unique minimizer lies on the positive branch, the negative branch, or at \(v_k=0\).
The one-sided KCL residuals at the switching point between the two branches are
\begin{equation}
F_+(0)=b_k+I_s\exp\!\left(-\frac{v_{\mathrm{off},1}}{v_T}\right),
\qquad
F_-(0)=b_k-I_s\exp\!\left(-\frac{v_{\mathrm{off},2}}{v_T}\right).
\label{eq:double_diode_one_sided_residuals}
\end{equation}  

The active branch is selected from the signs of these one-sided residuals:
\begin{equation}
\begin{aligned}
F_+(0)<0
&\quad\Longrightarrow\quad
v_k^\star>0,\\[2pt]
F_-(0)>0
&\quad\Longrightarrow\quad
v_k^\star<0,\\[2pt]
F_-(0)\le 0 \le F_+(0)
&\quad\Longrightarrow\quad
v_k^\star=0.
\end{aligned}
\label{eq:double_diode_branch_selection}
\end{equation}

When the minimizer lies on a nonzero branch, its value is easily obtained by rewriting the stationarity condition as a Lambert-W equation. Using Eq.~\eqref{eq:double_diode_large_signal}, the stationarity equation reduces to a single-exponential equation:
  \begin{equation}
  A\,x+B+C\,e^{D x}=0,
  \qquad A\neq 0,\; D\neq 0.
  \label{eq:double_diode_lambert_generic_form}
  \end{equation}
  \textbf{Case (i). }
On the positive branch
the solution is given by 
  \begin{equation}
  v_k^\star
  =
  -\frac{b_k}{2a_k}
  -\;v_T\,W_0\!\left(
  \frac{I_s}{2a_k v_T}\,
  \exp\!\Big(
  -\frac{v_{\mathrm{off},1}}{v_T}
  -\frac{b_k}{2a_k v_T}
  \Big)
  \right),
  \qquad v_k^\star>0.
  \label{eq:double_diode_LS_vstar_pos}
  \end{equation}
\\
  \textbf{Case (ii). }
  On the negative branch the solution is given by
  \begin{equation}
  v_k^\star
  =
  -\frac{b_k}{2a_k}
  +\;v_T\,W_0\!\left(
  \frac{I_s}{2a_k v_T}\,
  \exp\!\Big(
  -\frac{v_{\mathrm{off},2}}{v_T}
  +\frac{b_k}{2a_k v_T}
  \Big)
  \right),
  \qquad v_k^\star<0.
  \label{eq:double_diode_LS_vstar_neg}
  \end{equation}
The full derivation is provided in Supplementary Section~\ref{sec:supp_double_shockley_lambert}.

\paragraph*{Piecewise-linear i-v curve.}
\label{subsec:exp_iv}

The Shockley nonlinearities above are specified by analytic \(i\)-\(v\) expressions. In practice, however, a nonlinear device may be available through measured or simulated tabulated \(i\)-\(v\) data. To make the CD solver applicable to such empirical device models, we represent the nonlinear branch by a monotone piecewise-linear (PWL) interpolant.

For each nonlinear element, we assume that a set of points
\(\{(v_m,i_m)\}_{m=0}^{M}\) is given, with
\[
v_0 < \dots < v_M,
\qquad
i_0 \le \dots \le i_M .
\]
From these points we construct a monotone PWL interpolant \(\hat{i}(v)\), so that on each interval \([v_m,v_{m+1}]\),
\begin{equation}
\hat{i}(v) = s_m v + c_m,
\qquad
v \in [v_m,v_{m+1}],
\label{eq:pwl_iv_segment}
\end{equation}
where
\begin{equation}
s_m = \frac{i_{m+1}-i_m}{v_{m+1}-v_m},
\qquad
c_m = i_m - s_m v_m.
\label{eq:pwl_slope_intercept}
\end{equation}

The stationary condition is then given by
\begin{equation}
\frac{\partial E}{\partial v_k}(v)
\;=\;
2a_k v_k + b_k + \hat{i}(v_k)
\;=\;0.
\label{eq:stationary_exp}
\end{equation}

Eq.~\eqref{eq:stationary_exp} is a one-dimensional root-finding problem, which we solve using the Newton--Raphson method.
Since \(\hat{i}(v)\) is piecewise linear, it is differentiable on each interval
\((v_m,v_{m+1})\), with
\[
\hat{i}'(v)=s_m
\qquad
\text{for } v\in(v_m,v_{m+1}).
\]
We therefore compute the coordinate update with a set of Newton iterations, using the slope of the active
PWL segment:
\begin{equation}
v_k^{(t+1)}
=
v_k^{(t)}
-
\frac{2a_k v_k^{(t)} + b_k + \hat{i}\!\bigl(v_k^{(t)}\bigr)}
     {2a_k + \hat{i}'\!\bigl(v_k^{(t)}\bigr)}.
\label{eq:newton_exp}
\end{equation} 
 The active PWL segment is updated after each step, and the process is repeated until convergence.

 \subsection*{Application to deep resistive networks}

\paragraph*{CD updates.}

The networks considered above have arbitrary architecture, where nodes may be connected by any resistor branches and nonlinear elements are
attached to selected nodes. Deep resistive networks (DRNs) \cite{Kendall2020} are a
subclass of these circuits, designed to mirror the layered architecture of conventional neural networks. The nodes
are partitioned into layers \(0,\dots,L\), with layer \(0\) corresponding to the input, layer \(L\) to the
output, and intermediate layers to hidden units. The trainable parameters of the system are the conductances \(g^{(\ell)}_{jk}\), which connect only adjacent layers. Nonlinear elements are attached to hidden-layer nodes, while the output
layer is linear.

The CD framework developed above applies directly to DRNs:
the same scalar CD update is used at each node and the layered structure gives explicit formulas for
the local coefficients \(a_k^{(\ell)}\) and \(b_k^{(\ell)}\). 
The energy function of the DRN can be written in a layerwise fashion: 
\begin{equation}
E(v)
\;=\;
\frac{1}{2}\sum_{\ell=1}^{L}
\sum_{j=1}^{N_{\ell-1}}\sum_{k=1}^{N_{\ell}}
g^{(\ell)}_{jk}\,\bigl(v^{(\ell-1)}_{j}-v^{(\ell)}_{k}\bigr)^2
\;+\;
\sum_{\ell=1}^{L-1}
\sum_{k=1}^{N_{\ell}}
\phi^{(\ell)}_{k}\!\bigl(v^{(\ell)}_{k}\bigr).
\label{eq:drn_energy_with_nl_vk}
\end{equation}

The Methods section ``CD equations for deep resistive networks'' derives the expression of CD updates in this case. As suggested in \cite{Scellier2024}, the layered structure makes the CD updates parallelizable.
As there are no conductive couplings within a layer, all nodes in a layer can be updated independently once the neighboring layers are fixed.
In networks with many layers, one can alternate between updating all odd-indexed layers while keeping the even layers fixed, and updating all even-indexed layers while keeping the odd layers fixed. Each half-iteration can then be written in matrix-vector form, which makes the solver efficient on multicore CPUs or GPUs.

In deep DRNs, voltage signals attenuate as they propagate through successive layers. To mitigate this signal attenuation while  preserving the network's bidirectionality, ref.~\cite{Kendall2020}
introduced a bidirectional amplifier between adjacent layers. The amplifier consists of a
voltage-controlled voltage source (VCVS), which amplifies the forward voltage
by a factor \(A\), and a current-controlled current source (CCCS), which amplifies
the backward current by a factor \(B\). In ref.~\cite{Kendall2020}, the
reciprocal choice \(B=1/A\) was used. Here, we allow $A$ and $B$ to be chosen independently and
show how the resulting rescaling can be incorporated into the energy function of the network and the CD framework. The Methods section ``Bidirectional amplification'' derives the exact CD updates in the presence of such arbitrary bidirectional amplification.

%\newpage
\paragraph*{SPICE validation.}

\begin{figure}[!htbp]
    \centering
    \includegraphics[width=\linewidth]{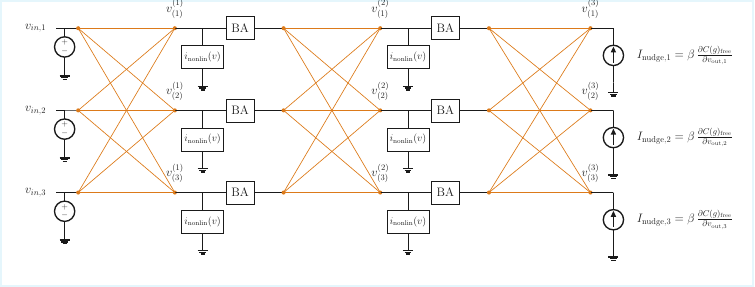}
    \caption{\textbf{Architecture of the DRN used in the experiments.} Dense  conductance layers are connected by grounded nonlinearities and bidirectional amplifiers. In both the free and nudged phases, the input nodes are clamped to the input voltages. In the nudged phase, the nudging currents are additionally injected into the output nodes.}
    \label{fig:deep_resistive_network}
\end{figure}

We implement the proposed CD solver in PyTorch and evaluate it against matched SPICE simulations of the same DRNs (see Methods). In the CD-SPICE comparisons, the two simulators use the same conductance matrices, input clamps, output current sources, nonlinear branch laws, and bidirectional amplification parameters. Fig.~\ref{fig:deep_resistive_network} shows the DRN architecture used in the experiments. Unless otherwise stated, all simulations use bidirectional amplification factors $A=4$ and $B=1$. The voltage amplification helps compensate signal attenuation across layers. We use $B=1$, rather than the reciprocal choice $B=1/A$ used in ref.~\citet{Kendall2020}, because reducing the backward current gain made deeper DRNs harder to train in preliminary experiments.

We first evaluate whether CD reproduces the steady states obtained by SPICE.
All results in this subsection use DRNs trained on the Digits dataset and are evaluated on the same 360-sample validation set in both simulators. We consider one-, two-, and three-hidden-layer architectures,  hidden layer widths from 64 to 1024 and the three nonlinearities studied in this work: a Shockley diode, a double-Shockley diode, and a PWL \(i(v)\) curve. 
The single-diode networks use output dimension 10, while the double-diode and PWL networks use output dimension 20.
SPICE could not generate netlists for the architectures with two hidden layers and hidden width 1024, nor for those with three hidden layers and hidden widths 512 or 1024. These configurations are therefore excluded from the matched CD-SPICE comparisons.

The PWL nonlinearity is included to test the solver on a tabulated device law. Here it is constructed synthetically from the double-Shockley curve in Eq.~\eqref{eq:double_diode}: we sample the curve at 200 uniformly spaced points on $[-2,2]$~V and linearly interpolate between the samples to obtain
$\hat{i}(v)$.

%For the single- and double-Shockley diode runs, we use over-relaxed CD updates in the deeper architectures. After computing the exact coordinate update, the next iterate is displaced toward that update with over-relaxation factor \(\omega=1.2\) for two-hidden-layer networks and \(\omega=1.4\) for three-hidden-layer networks. The implementation and effect of over-relaxation are described in Supplementary Sec.~\ref{sec:supp_timing_analysis}.
For the single- and double-Shockley diode runs, we use an over-relaxed variant of coordinate descent in the deeper architectures to accelerate convergence. After computing the exact coordinate update \(v_k^{\mathrm{CD}}\), the node voltage is updated according to
\[
v_k^{\mathrm{new}} = v_k^{\mathrm{old}} + \omega\left(v_k^{\mathrm{CD}} - v_k^{\mathrm{old}}\right).
\]
The standard CD update is recovered for \(\omega=1\), whereas \(\omega>1\) extrapolates beyond the exact coordinate update in the same direction. We use \(\omega=1.2\) for two-hidden-layer networks and \(\omega=1.4\) for three-hidden-layer networks. The implementation and effect of over-relaxation are described in Supplementary Sec.~\ref{sec:supp_timing_analysis}.

%For the single- and double-Shockley diode runs, we use over-relaxed CD steps in the deeper architectures.
%After computing the exact coordinate update, the next iterate is displaced toward that update with over-relaxation factor $\omega=1.2$ for two-hidden-layer networks and $\omega=1.4$ for three-hidden-layer networks.  The implementation and effect of the over-relaxed update is detailed in the Supplementary Sec. \ref{sec:supp_timing_analysis}.

To quantify the agreement between CD and SPICE, we compare the steady-state node voltages. 
For a validation sample $i$, let \(\mathbf{v}^{\mathrm{CD}}(i)\) and \(\mathbf{v}^{\mathrm{SPICE}}(i)\) denote the vectors obtained by stacking the hidden- and output-layer voltages.
We define the samplewise relative error as
\begin{equation}
  \mathrm{RelErr}_i
  \;=\;
  \frac{\left\|\mathbf{v}^{\mathrm{CD}}(i)-\mathbf{v}^{\mathrm{SPICE}}(i)\right\|_{1}}
       {\left\|\mathbf{v}^{\mathrm{SPICE}}(i)\right\|_{1}}
  \;=\;
  \frac{\sum_{k=1}^{N}\left|v_k^{\mathrm{CD}}(i)-v_k^{\mathrm{SPICE}}(i)\right|}
       {\sum_{k=1}^{N}\left|v_k^{\mathrm{SPICE}}(i)\right|}.
\end{equation}

For each configuration, we summarize \(\{\mathrm{RelErr}_i\}_{i=1}^{360}\) by its 90th percentile (P90).

%We first run CD for a fixed number of sweeps and compare the resulting voltages with SPICE. Here, one CD sweep denotes one update of all odd-indexed layers followed by one update of all even-indexed layers. For each nonlinearity and each depth, we evaluate a width-\(128\) Digits network and report the P90 relative \(L_1\) error over the validation set.

We define a CD sweep as one full pass through the non-clamped nodes of the network. In the odd-even layer update scheme used here, this corresponds to updating all odd-indexed layers, followed by all even-indexed layers, so that each free node is updated once per sweep. We then evaluate how the CD solution approaches the SPICE equilibrium as the number of sweeps increases. For each nonlinearity and each depth, we use a width-128 Digits network. We summarize the validation error by its 90th percentile, meaning that 90\% of validation samples have a relative $L_1$ voltage error below the reported value.

\begin{figure}[!htbp]
    \thisfloatpagestyle{empty}
    \centering

    \begin{subfigure}[t]{0.70\linewidth}
        \centering
        \includegraphics[width=\linewidth]{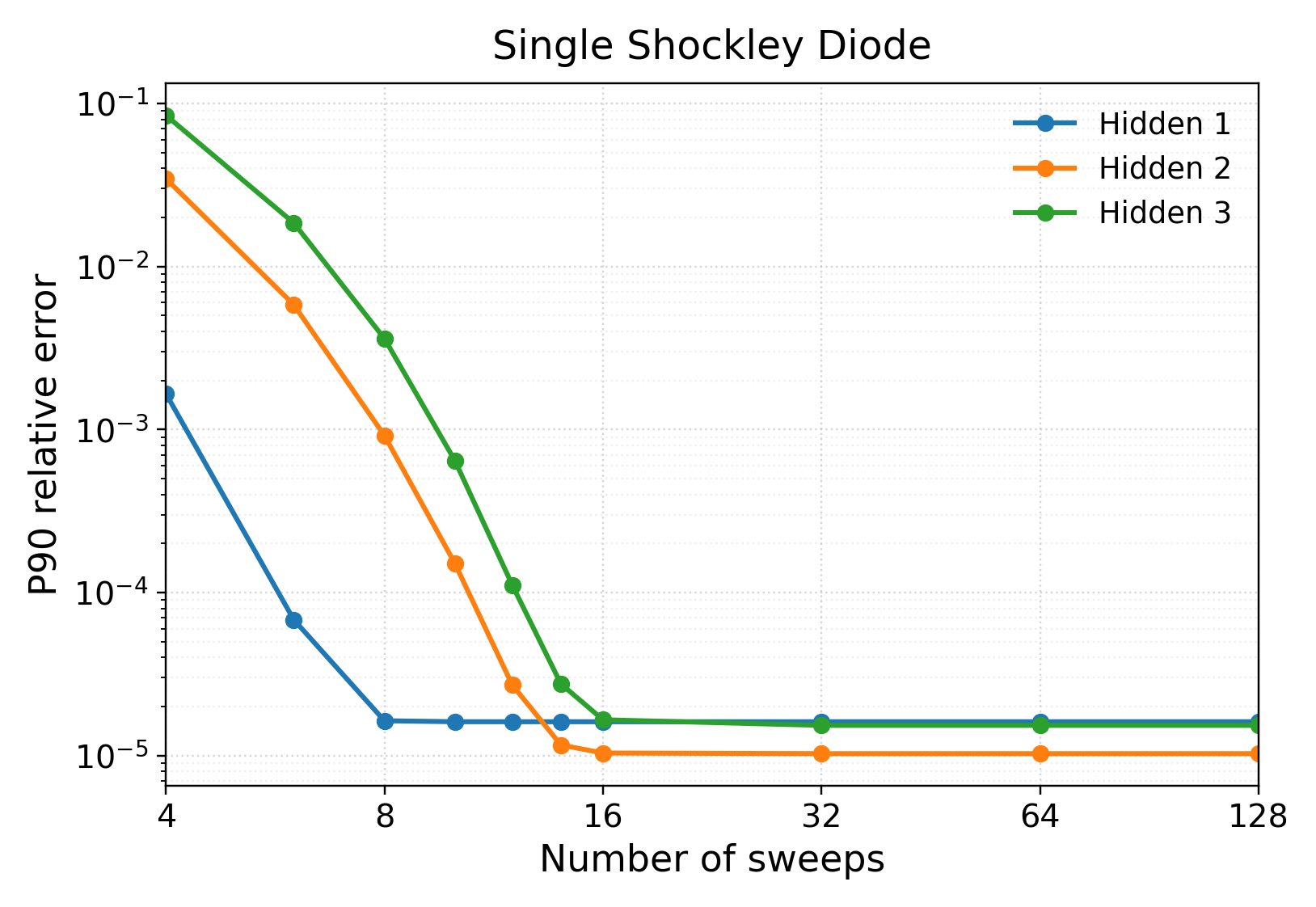}
        \caption{Single Shockley Diode.}
        \label{fig:error_vs_iterations_single_shockley}
    \end{subfigure}

    \vspace{0.8em}

    \begin{subfigure}[t]{0.70\linewidth}
        \centering
        \includegraphics[width=\linewidth]{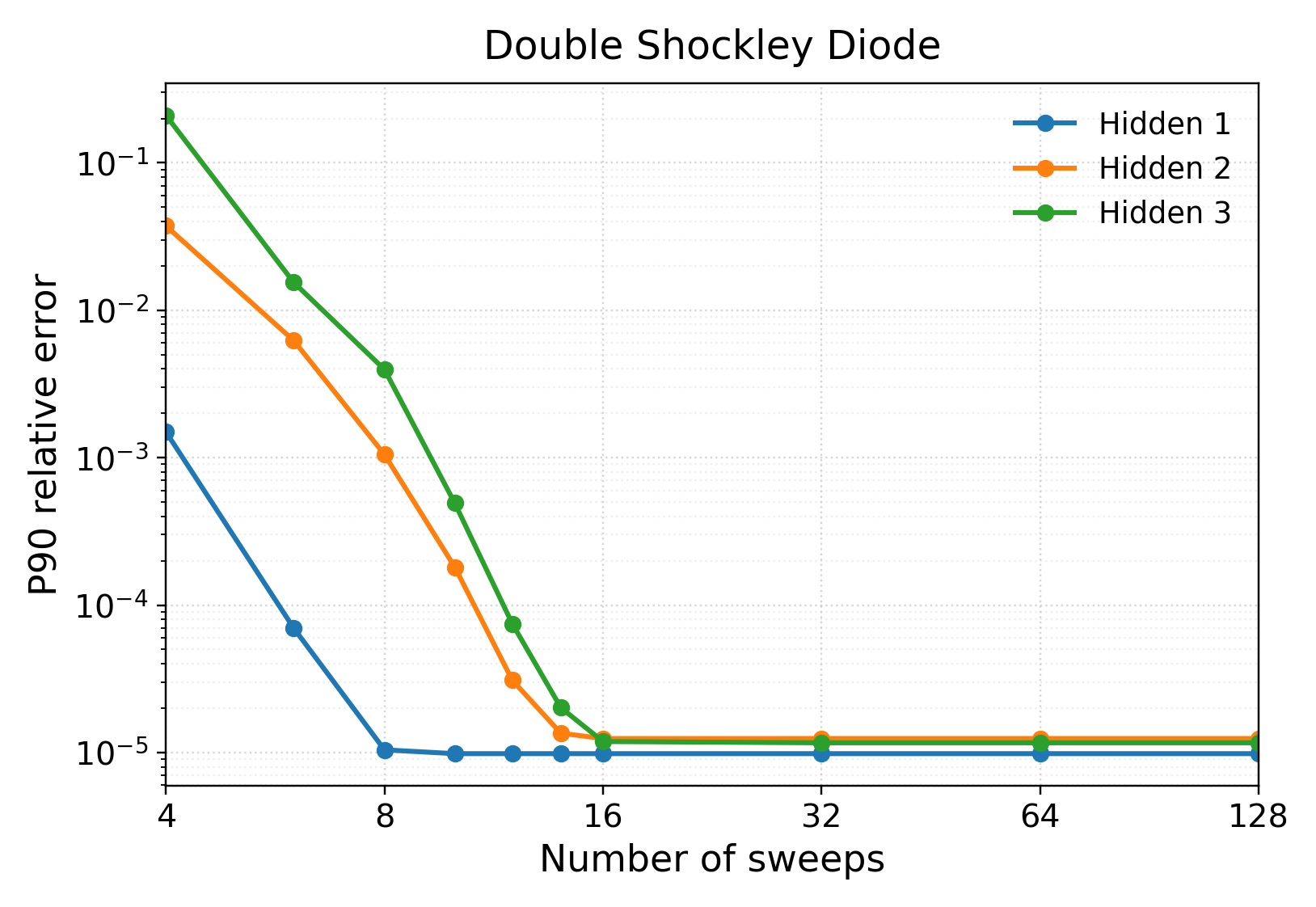}
        \caption{Double Shockley Diode.}
        \label{fig:error_vs_iterations_double_shockley}
    \end{subfigure}

    \vspace{0.8em}

    \begin{subfigure}[t]{0.70\linewidth}
        \centering
        \includegraphics[width=\linewidth]{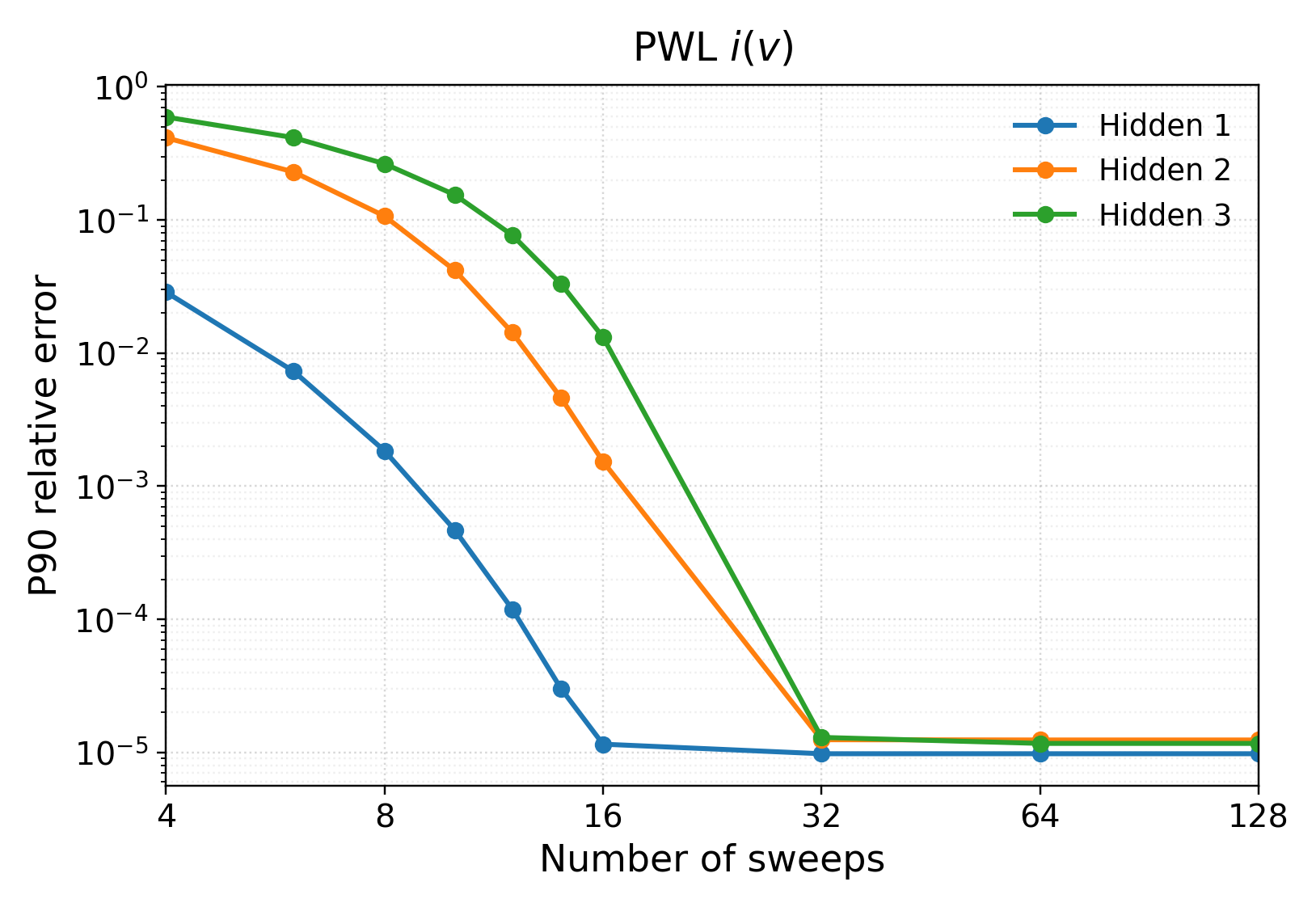}
        \caption{PWL \(i(v)\).}
        \label{fig:error_vs_iterations_experimental}
    \end{subfigure}

    \caption{P90 relative \(L_1\) voltage error between CD and SPICE versus the number of CD sweeps for width-\(128\) Digits networks. Each panel shows one nonlinearity. Hidden 1, Hidden 2 and Hidden 3 denote  networks with one, two, and three hidden layers.}
    \label{fig:error_vs_iterations}
\end{figure}

Fig.~\ref{fig:error_vs_iterations} shows that, across all network depths and nonlinearities, the error decreases rapidly during the first few sweeps and then saturates at a low floor. Across all reported configurations, the final P90 relative error is of order $10^{-5}$. Increasing the number of hidden layers shifts the curves to the right, indicating that deeper networks require more sweeps to reach the same accuracy. All networks with the single and double Shockley nonlinearities reach their observed error floors within 16 sweeps, while those with the PWL nonlinearity require up to 32 sweeps to converge.

\subsection*{Runtime scaling}

The fixed-sweep experiments in the previous section show that CD approaches the SPICE steady state, but a fixed number of sweeps is not ideal for timing comparisons because the number of sweeps needed for convergence depends on the conductances, nonlinearity, width, and depth. For the runtime experiments, we therefore use an adaptive stopping rule based on the relative voltage change. Writing $\mathbf{v}^{(t)}$ for the stacked vector of all non-clamped node voltages after sweep $t$, convergence is declared when
\begin{equation}
\|\mathbf{v}^{(t+1)}-\mathbf{v}^{(t)}\|_{\infty}
\;\le\;
v_{\mathrm{tol}}\,\|\mathbf{v}^{(t+1)}\|_{\infty} + v_{n,\mathrm{tol}},
\end{equation}
where \(v_{n,\mathrm{tol}}=10^{-6}\,\mathrm{V}\) is a small absolute floor.

\begin{figure}[!htbp]
    \thisfloatpagestyle{empty}
    \centering

    \begin{subfigure}[t]{0.70\linewidth}
        \centering
        \includegraphics[width=\linewidth]{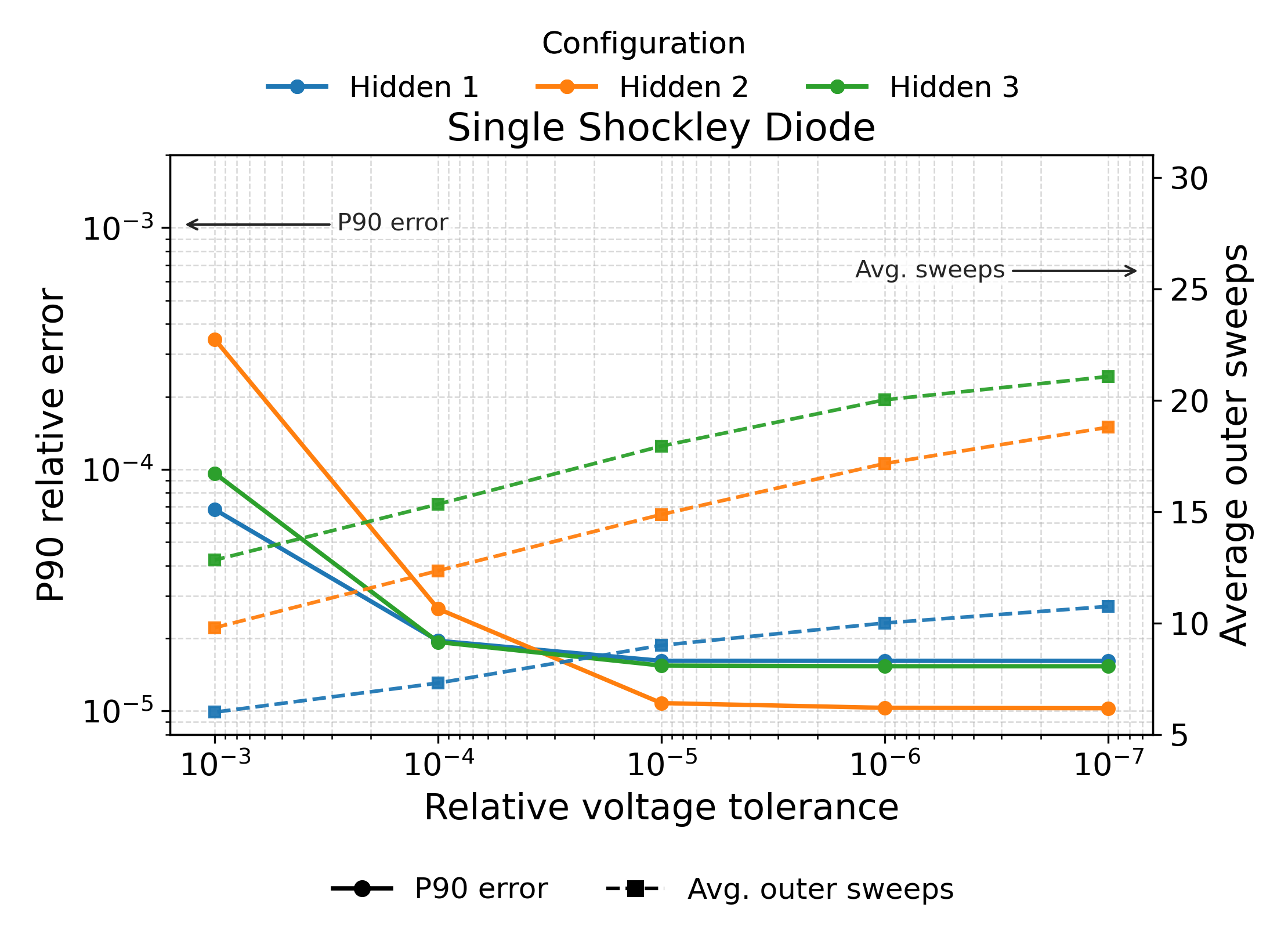}
        \caption{Single Shockley Diode.}
        \label{fig:error_vs_voltage_tolerance_single_shockley}
    \end{subfigure}

    \vspace{0.35em}

    \begin{subfigure}[t]{0.70\linewidth}
        \centering
        \includegraphics[width=\linewidth]{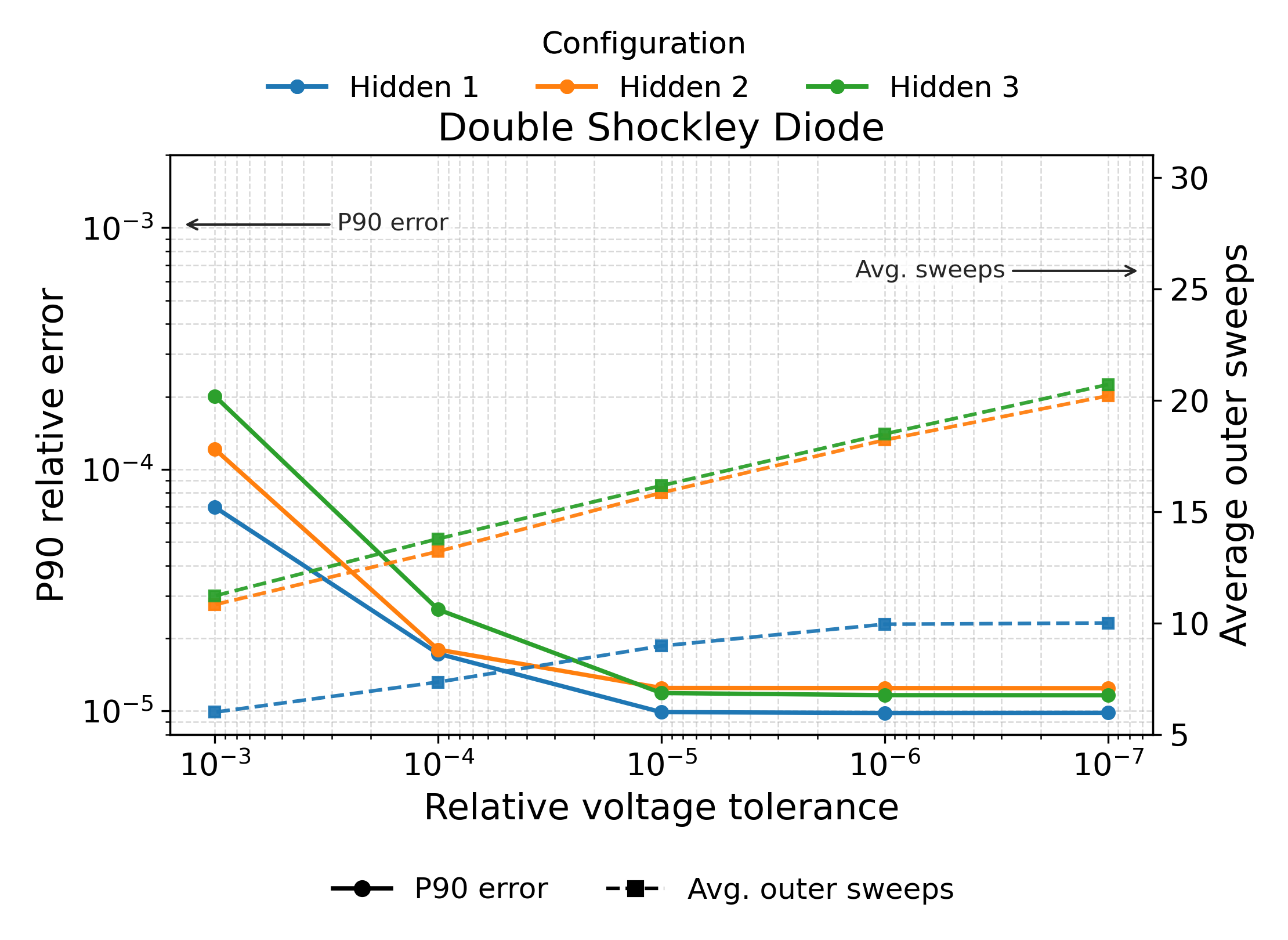}
        \caption{Double Shockley Diode.}
        \label{fig:error_vs_voltage_tolerance_double_shockley}
    \end{subfigure}

    \vspace{0.35em}

    \begin{subfigure}[t]{0.70\linewidth}
        \centering
        \includegraphics[width=\linewidth]{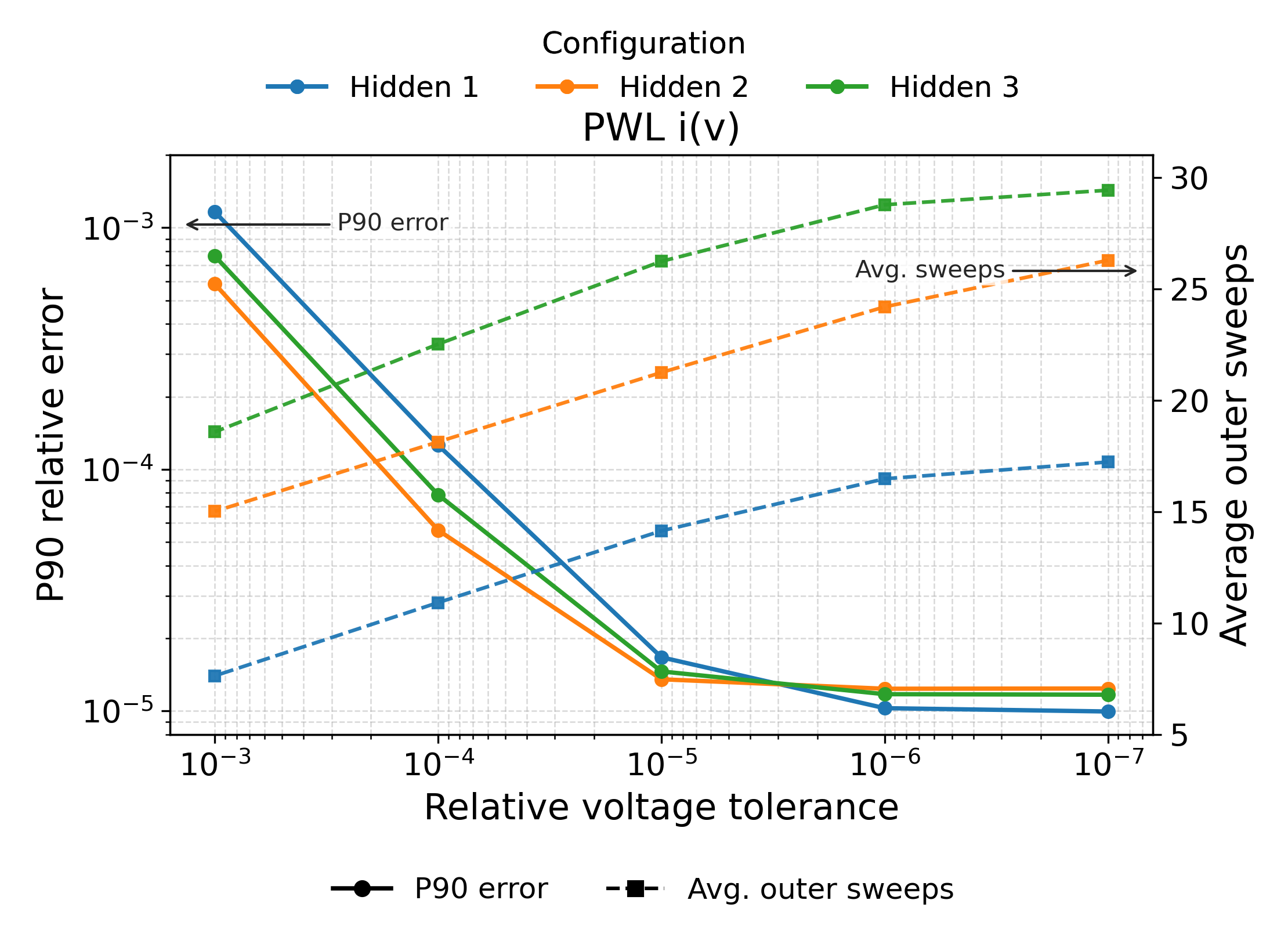}
        \caption{PWL \(i(v)\).}
        \label{fig:error_vs_voltage_tolerance_experimental}
    \end{subfigure}

    \caption{ P90 relative \(L_1\) voltage error and average outer sweeps versus the relative-voltage tolerance \(v_{\mathrm{tol}}\) for width-\(128\) Digits networks. Each panel shows one nonlinearity. Hidden 1, Hidden 2 and Hidden 3 denote  networks with one, two, and three hidden layers. Solid curves show P90 errors on the left axis and dashed curves show average outer sweeps on the right axis.}
    \label{fig:error_vs_voltage_tolerance}
\end{figure}

We evaluate this stopping rule on the same width-128 Digits networks
used above by varying the relative voltage tolerance
\(v_{\mathrm{tol}}\). Fig.~\ref{fig:error_vs_voltage_tolerance} shows
how the P90 relative \(L_1\) CD--SPICE error and the average number of
sweeps depend on this tolerance. Across the three nonlinearities,
tightening \(v_{\mathrm{tol}}\) reduces the P90 error, but the
improvement becomes marginal once \(v_{\mathrm{tol}}\) reaches
\(10^{-5}\), where the error is already on the order of \(10^{-5}\). The tolerance, 
\(v_{\mathrm{tol}}=10^{-5}\) provides a good accuracy--cost trade-off as 
tighter tolerances require additional sweeps while providing little 
further improvement in error. We therefore use this value in the 
main runtime experiments.

We next compare the wall-clock time of CD and SPICE on the full Digits validation set. For each choice of nonlinearity, depth, and hidden-layer width, we evaluate the corresponding trained model with both solvers. Both CD and SPICE measurements are performed on CPU, on a dual-socket machine with Intel Xeon Gold 5220 CPUs. To measure serial equilibrium-solve cost rather than batching performance, the CD runs use the batch size of one. CD is run with the adaptive stopping rule above, using \(v_{\mathrm{tol}}=10^{-5}\). Reported SPICE timings correspond to DC simulation time and exclude netlist generation.

%CD is faster than SPICE for every configuration with a reliable matched SPICE timing. Supplementary Tables~\ref{tab:timing_runtime_decomposition} and ~\ref{tab:timing_accuracy_summary} show speedups ranging from \(1.2\times\) to \(1750\times\), while agreement with SPICE remains tight: the P90 relative \(L_1\) error stays below \(1.0\times 10^{-4}\) and is typically in the \(10^{-5}\) range.

\begin{figure}[!htbp]
    \thisfloatpagestyle{empty}
    \centering

    \begin{subfigure}[t]{0.74\linewidth}
        \centering
        \includegraphics[width=\linewidth]{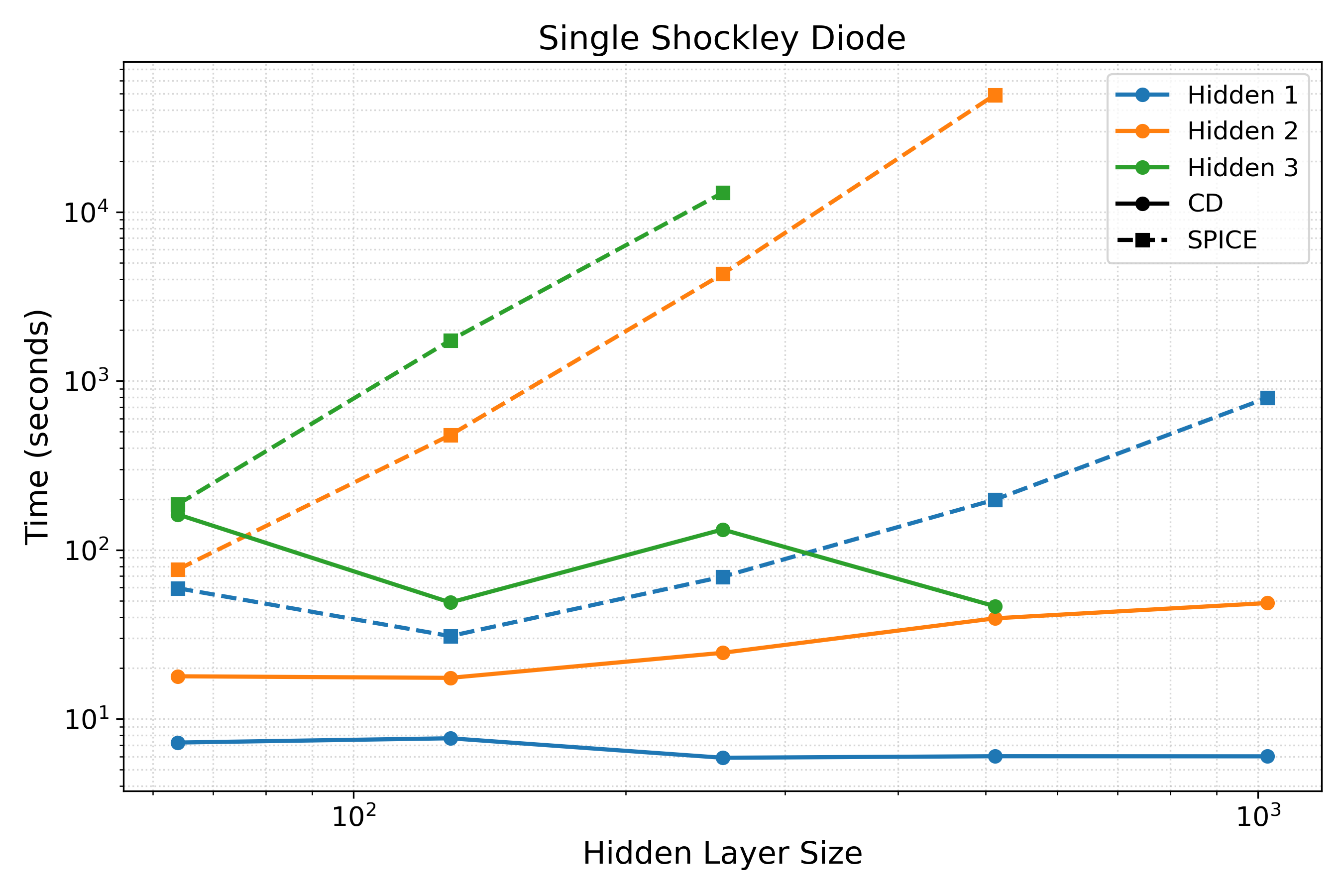}
        \caption{Single Shockley Diode.}
        \label{fig:runtime_vs_size_single_cpu}
    \end{subfigure}

    \vspace{0.8em}

    \begin{subfigure}[t]{0.74\linewidth}
        \centering
        \includegraphics[width=\linewidth]{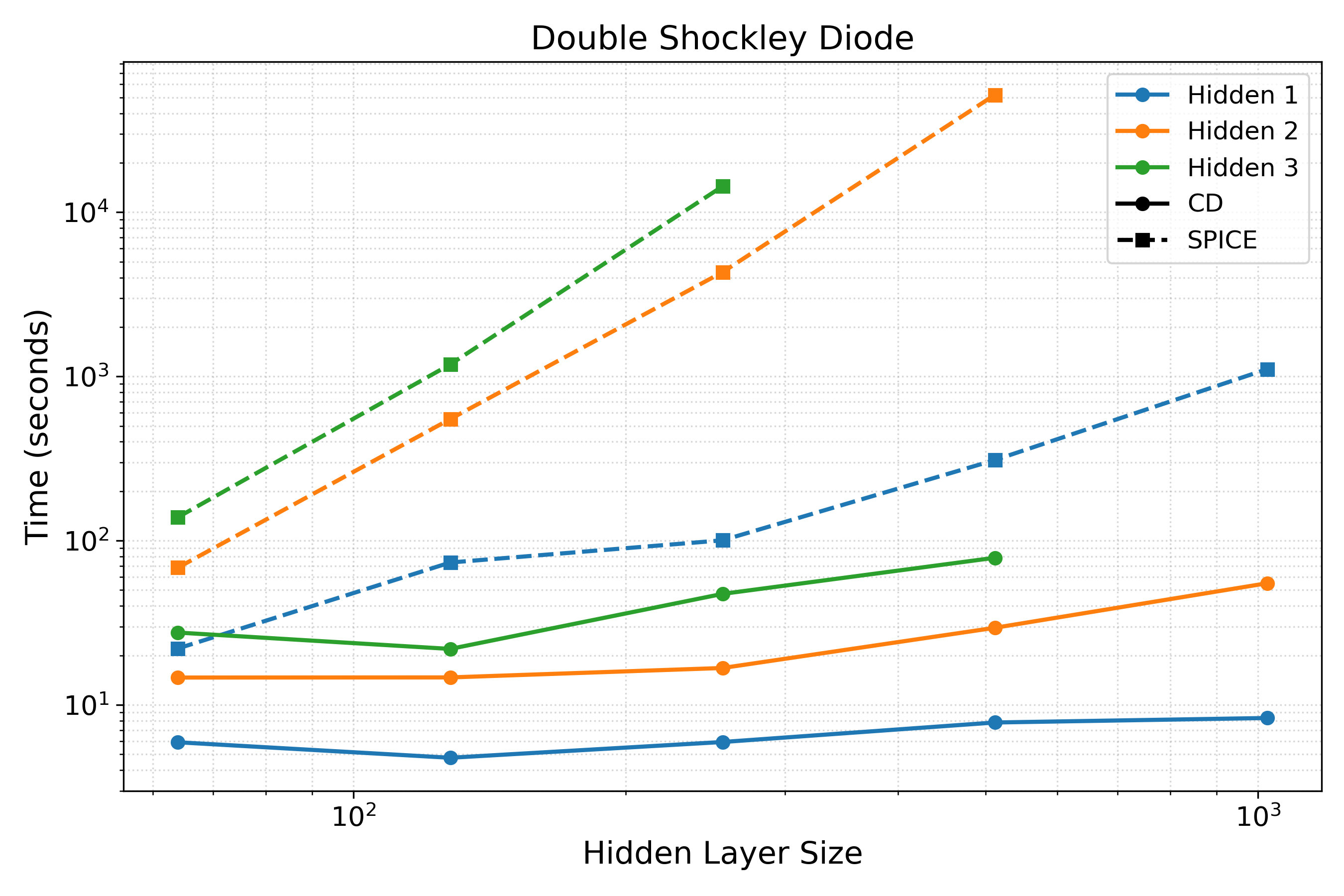}
        \caption{Double Shockley Diode.}
        \label{fig:runtime_vs_size_double_cpu}
    \end{subfigure}

    \vspace{0.8em}

    \begin{subfigure}[t]{0.74\linewidth}
        \centering
        \includegraphics[width=\linewidth]{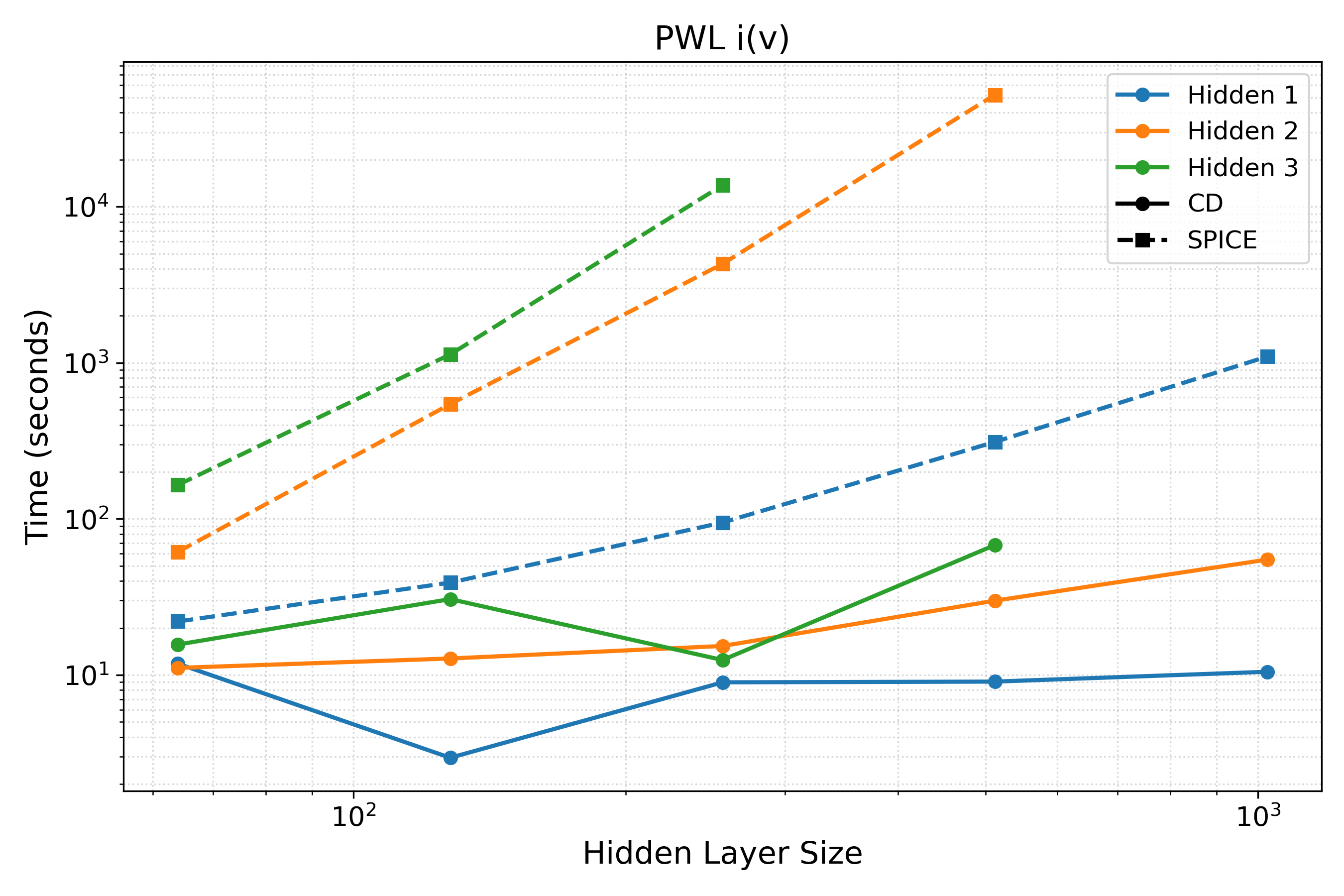}
        \caption{PWL \(i(v)\).}
        \label{fig:runtime_vs_size_experimental_cpu}
    \end{subfigure}

    \caption{Runtime versus hidden width for matched CD and SPICE simulations. Each panel shows one nonlinearity. Hidden 1, Hidden 2 and Hidden 3 denote  networks with one, two, and three hidden layers. Solid curves show CD runtimes and dashed curves show SPICE runtimes.}
    \label{fig:runtime_vs_size}
\end{figure}

Fig.~\ref{fig:runtime_vs_size} shows the runtime scaling. SPICE runtime rises steeply with width and depth, while CD has more moderate scaling. The CD runtime is controlled by two factors: the number of outer odd-even sweeps needed to satisfy the stopping rule, and the cost of the local nonlinear updates within each sweep. For the exponential nonlinearities, the scalar update has a closed-form Lambert-\(W\) expression. In the implementation, extreme Lambert arguments are clipped for numerical stability, and the resulting local KCL residual is refined by Newton-Raphson iterations when needed. Supplementary Sec.~\ref{sec:supp_timing_analysis} analyzes this runtime decomposition in more detail.

CD is faster than SPICE for every configuration with a reliable matched SPICE timing. Supplementary Tables~\ref{tab:timing_runtime_decomposition} and ~\ref{tab:timing_accuracy_summary} show speedups ranging from \(1.2\times\) to \(1750\times\), while agreement with SPICE remains tight: the P90 relative \(L_1\) error stays below \(1.0\times 10^{-4}\) and is typically in the \(10^{-5}\) range.

\subsection*{MNIST-scale training}

Finally, we test whether CD can be used to train a larger DRN. We train a \(1568\times100\times20\) network with double antiparallel Shockley-diode nonlinearities on MNIST. This is the same DRN size and device family studied
by \citet{Kendall2020}, but trained here with the CD solver.

For all MNIST experiments, we use a fixed budget of four CD sweeps per equilibrium solve.  A fixed sweep budget is more convenient for batched training than an adaptive stopping rule, and increasing the budget to six or eight sweeps did not produce a meaningful improvement in final test error.

\begin{figure}[!htbp]
    \centering
    \begin{subfigure}[t]{0.76\linewidth}
        \centering
        \includegraphics[width=\linewidth]{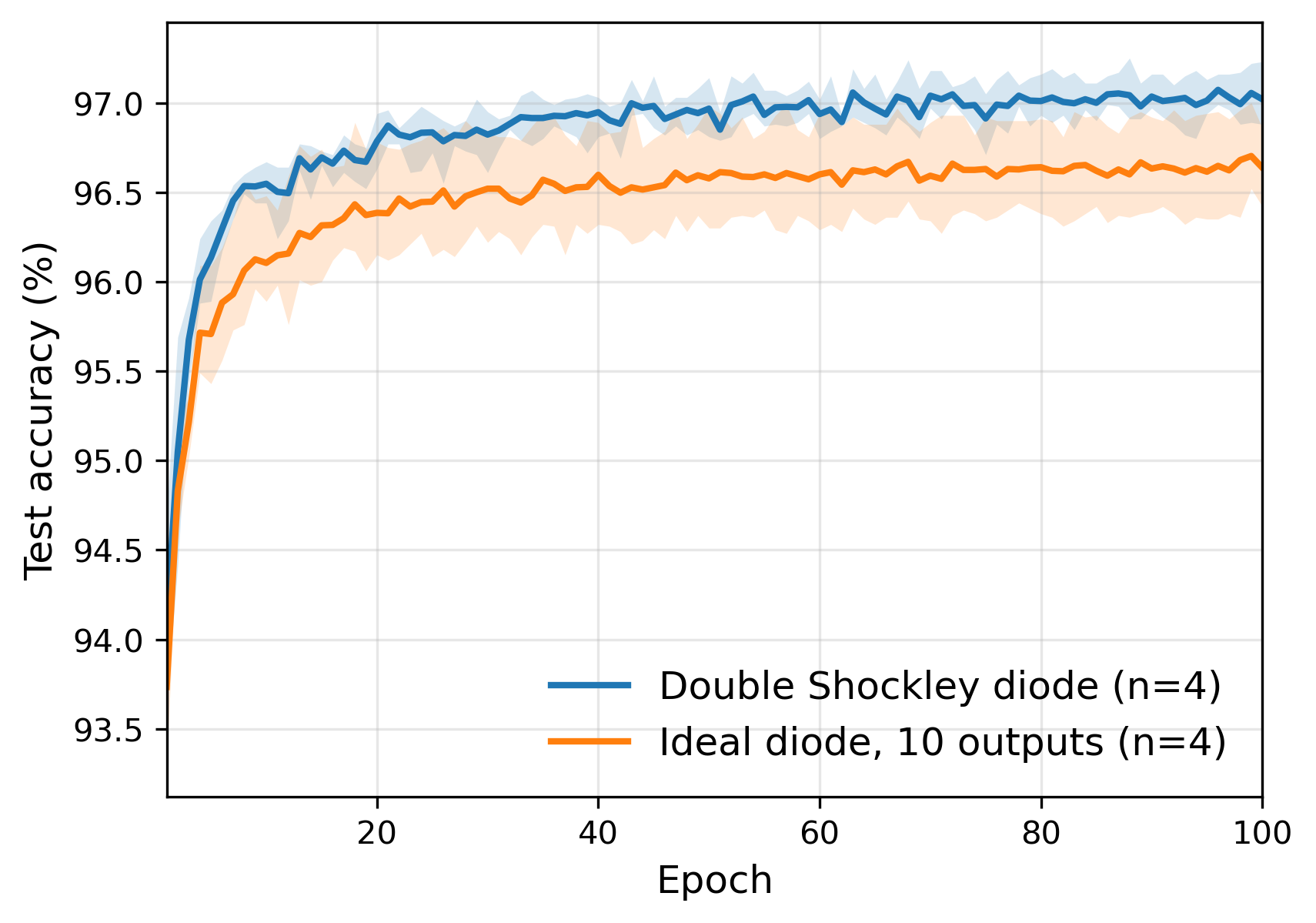}
        \caption{MNIST test accuracy.}
        \label{fig:mnist_mean_accuracy}
    \end{subfigure}

    \vspace{0.8em}

    \begin{subfigure}[t]{0.92\linewidth}
        \centering
        \includegraphics[width=\linewidth]{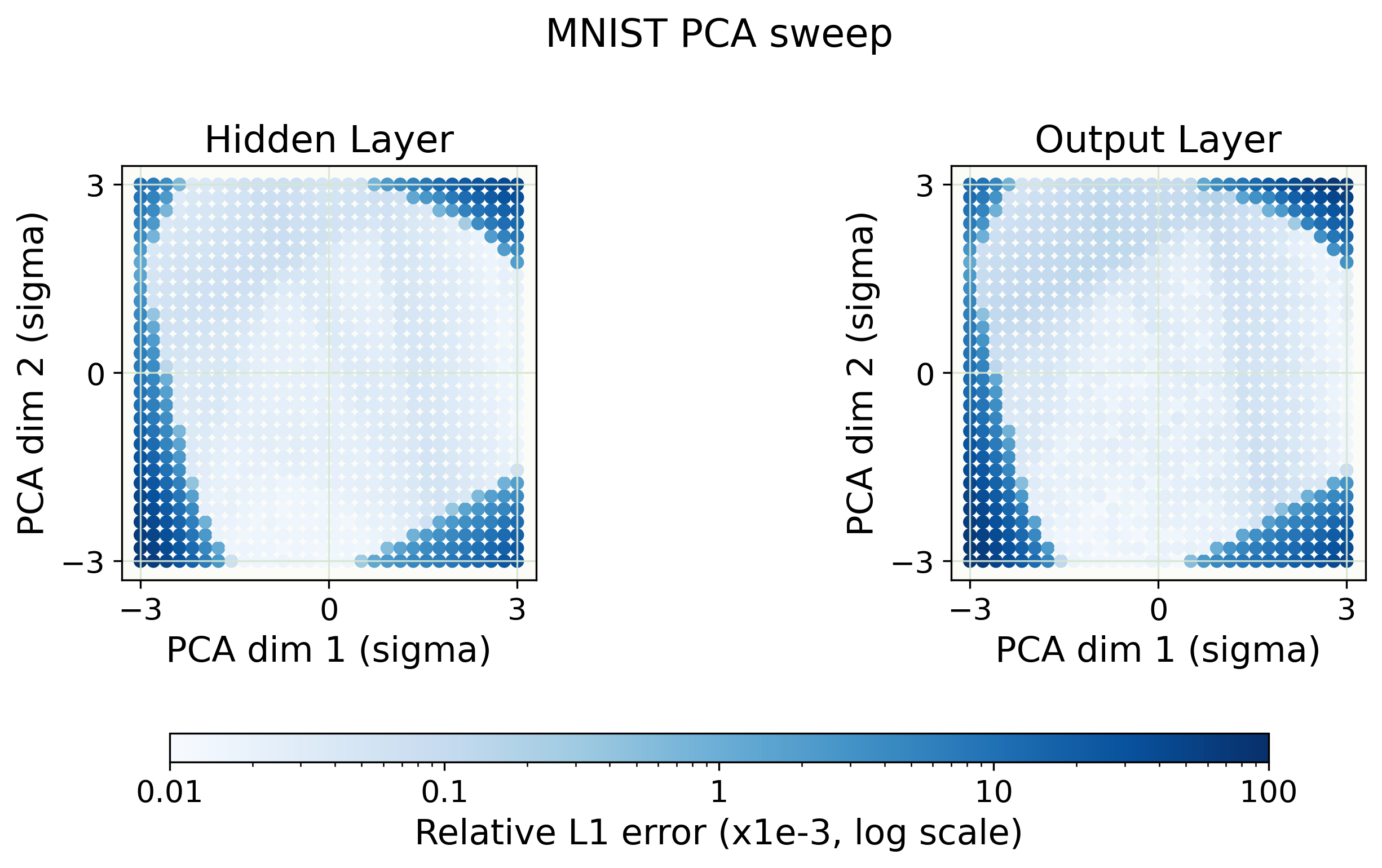}
        \caption{CD--SPICE PCA-sweep error.}
        \label{fig:pca_error}
    \end{subfigure}
    \caption{MNIST-scale simulations. Top: test accuracy across four independent 100-epoch training runs of the \(1568\times100\times20\) network with the double Shockley diode, together with the \(1568\times100\times10\) ideal-diode baseline from Ref.~\cite{Scellier2024}. All runs use four CD sweeps. The central curves show the mean across runs and the shaded bands show the min--max envelopes. Bottom: P90 relative \(L_1\) error between CD and SPICE across a \(30\times30\) PCA sweep of the MNIST input space for the trained \(1568\times100\times20\) network.}
    \label{fig:mnist_accuracy_and_pca_error}
\end{figure}

Fig.~\ref{fig:mnist_mean_accuracy} summarizes the test accuracy across four independent double-diode runs and includes the output-dimension-10 ideal-diode baseline for comparison. The final test error of the double-diode runs is around \(3\%\) and the full training run takes roughly 3.5 hours in our implementation. By comparison, ~\cite{Kendall2020} reported about \(3.4\%\) test error after 10 training epochs and roughly one week of training for the same network class. Thus,
CD enables substantially longer training and yields slightly better final accuracy. After training, we exported the learned weights to SPICE, and verified that the same accuracy was recovered.
 
%To assess agreement between the two solvers we also compare CD and SPICE on a two-dimensional PCA slice of the input space. We use the first two principal directions and sweep a \(30\times30\) grid in the corresponding plane centered at the data mean, \[ v(\alpha,\beta)=\mu+\alpha q_1+\beta q_2, \] with \(\alpha\in[-3\sigma_1,\,3\sigma_1]\) and \(\beta\in[-3\sigma_2,\,3\sigma_2]\), where \(\sigma_i=\sqrt{\lambda_i}\). At each grid point, we compute the relative \(L_1\) error between the CD and SPICE steady states. Figure~\ref{fig:pca_error} reports the resulting P90 error over the sweep.

To assess agreement between the two solvers beyond the test samples, we also compare CD and SPICE on a two-dimensional PCA slice of the MNIST input space. We use the first two principal directions and sweep a \(30\times30\) grid in the corresponding plane centered at the data mean, \[ v(\alpha,\beta)=\mu+\alpha q_1+\beta q_2, \] with \(\alpha\in[-3\sigma_1,\,3\sigma_1]\) and \(\beta\in[-3\sigma_2,\,3\sigma_2]\), where \(\sigma_i=\sqrt{\lambda_i}\). At each grid point, we compute the relative \(L_1\) error between the CD and SPICE steady-state voltages, separately for the hidden and output layers. Figure~\ref{fig:pca_error} shows that the CD--SPICE discrepancy is smallest near the center of the PCA plane, where inputs remain close to the MNIST data distribution, and increases mainly near the edges and corners of the sweep. This indicates that the trained CD model remains consistent with the corresponding SPICE circuit over the relevant input region, with larger deviations confined mostly to out-of-distribution inputs.

\section*{Discussion}

%This observation is important for energy-based physical computation because practical analog circuits are built from non-ideal device laws. Shockley diodes, antiparallel diode pairs, and tabulated i-v curves are not naturally represented as ideal inequality constraints. Treating them through their dissipative content provides a device-aware route to simulation and training while preserving a direct connection to Kirchhoff's laws. The close agreement with matched SPICE simulations indicates that the proposed method solves the same circuit equilibrium problem to useful accuracy, rather than merely producing a qualitatively similar neural-network model.

This work extends fast coordinate-descent simulation of deep resistive learning machines from ideal-diode models to circuits with realistic nonlinear device characteristics. The key point is that, for grounded voltage-controlled nonlinear devices whose current-voltage characteristics are monotone, the electrical content remains convex and each coordinate update remains local. With neighboring voltages fixed, updating a node voltage reduces to enforcing Kirchhoff’s current law at that node. This allows realistic device behavior to be incorporated through the scalar node update, while preserving the layer-parallel odd-even structure that makes coordinate descent efficient in DRNs.

%The observed speedups change the practical scale of DRN studies. On Digits networks, CD remains within \(10^{-4}\) P90 relative \(L_1\) voltage error of SPICE while reducing runtime by up to \(1750\times\). On MNIST, the same approach makes it practical to train a \(1568\times100\times20\)  nonlinear DRN for 100 epochs and then export the trained circuit back to SPICE. The resulting classification accuracy is not intended as a digital machine-learning benchmark; rather, it demonstrates that realistic nonlinear resistive learning circuits can now be trained and validated at a scale where direct SPICE training is prohibitive.

The observed speedups change the practical scale of DRN studies. On Digits networks, CD reproduces SPICE steady-state voltages with relative \(L_1\) errors below \(10^{-4}\) for 90\% of validation samples, while reducing runtime by up to \(1750\times\). On MNIST, the same approach makes it practical to train a \(1568\times100\times20\) nonlinear DRN for 100 epochs and then export the trained circuit back to SPICE. The resulting classification accuracy is not intended as a digital machine-learning benchmark; rather, it demonstrates that realistic nonlinear resistive learning circuits can now be trained and validated at a scale where direct SPICE training is prohibitive.

Several idealizations remain. The present formulation assumes linear conductances, ideal sources, and one-port monotone nonlinearities. Extending the approach to nonlinear conductances, non-ideal sources, compact transistor models, and non-grounded nonlinear two-terminal elements would broaden its relevance to fabricated circuits. On the numerical side, the odd-even CD scheme could likely be accelerated further by Anderson acceleration, Krylov-like residual minimization, or multigrid-inspired corrections. Future work should also test whether these device-aware equilibrium solvers support deeper architectures and more challenging inference tasks.

\section*{Methods}

\subsection*{Simulation details}

The experiments using a fixed number of CD sweeps, and the experiments using a fixed voltage tolerance and the MNIST experiments were all performed on a machine equipped with an NVIDIA RTX 3090 GPU.

The Digits runtime comparison was performed on CPU on a dual-socket server equipped with Intel Xeon Gold 5220 processors. Both the CD and SPICE wall-clock measurements were obtained on this machine. To make the comparison with SPICE fair, the PyTorch CD simulations were run with batch size \(1\).

SPICE-level reference simulations were carried out with ELDO 17.1, a commercial SPICE-compatible circuit simulator. The runs were done using ELDO's interactive DC mode. We did not manually modify the simulator's accuracy settings; the default convergence parameters are reported in Table~\ref{tab:eldo_tolerances}.

\begin{table}[h!] \centering \caption{ELDO accuracy-control settings.} \label{tab:eldo_tolerances} \begin{tabular}{ll} \toprule Setting & Value \\ \midrule \texttt{RELTOL} & \(10^{-3}\) \\ \texttt{VNTOL} & \(10^{-6}\,\mathrm{V}\) \\ \texttt{ABSTOL} & \(10^{-12}\,\mathrm{A}\) \\ \texttt{ITOL} & \(10^{-6}\) \\ \texttt{EPS} & \(5\times10^{-3}\) \\ \bottomrule \end{tabular} \end{table}

\subsection*{CD equations for deep resistive networks}

The CD framework  applies directly to DRNs.
The energy function of the DRN can be written in a layerwise fashion: 
\begin{equation}
E(v)
\;=\;
\frac{1}{2}\sum_{\ell=1}^{L}
\sum_{j=1}^{N_{\ell-1}}\sum_{k=1}^{N_{\ell}}
g^{(\ell)}_{jk}\,\bigl(v^{(\ell-1)}_{j}-v^{(\ell)}_{k}\bigr)^2
\;+\;
\sum_{\ell=1}^{L-1}
\sum_{k=1}^{N_{\ell}}
\phi^{(\ell)}_{k}\!\bigl(v^{(\ell)}_{k}\bigr),
\label{eq:drn_energy_with_nl_vk}
\end{equation}
where the nonlinear dissipation potential on each branch is defined by
\begin{equation}
\phi^{(\ell)}_{k}(v)
\;:=\;
\int_{0}^{v} i^{(\ell)}_{k}(u)\,\mathrm{d}u.
\label{eq:phi_def_vk_drn}
\end{equation}
The nudging current sources are incorporated through the additional linear term
\[
E_{cs}(v)\;=\;\sum_{k=1}^{N_L} I^{(L)}_{k}\,v^{(L)}_{k},
\]
so that stationarity yields the source contribution
\[
\frac{\partial E_{cs}(v)}{\partial v^{(L)}_{k}}(v)\;=\; I^{(L)}_{k},
\qquad
(I^{(L)}_{k}=0\ \text{in the free phase}).
\]
The contribution of $v_k^{(\ell)}$ to the energy function of the network
can be written as
\begin{equation}
E_k^{(\ell)}\!\bigl(v_k^{(\ell)}\bigr)
\;=\;
a_k^{(\ell)}\bigl(v_k^{(\ell)}\bigr)^2
\;+\;
b_k^{(\ell)}\,v_k^{(\ell)}
\;+\;
\phi_k^{(\ell)}\!\bigl(v_k^{(\ell)}\bigr),
\label{eq:Ek_ab_phi_drn}
\end{equation}
where $\phi_k^{(\ell)}$ is the dissipation potential of the nonlinear device(s) attached to node $(\ell,k)$. \\
Here the coefficients $a_k$ and $b_k$ are defined as
\begin{equation}
a_k^{(\ell)}=
\begin{cases}
\frac{1}{2}\left(\displaystyle\sum_{j=1}^{N_{\ell-1}} g_{jk}^{(\ell)} \;+\; \sum_{m=1}^{N_{\ell+1}} g_{km}^{(\ell+1)}\right),
& \ell=1,\dots,L-1,\\[10pt]
\frac{1}{2}\displaystyle\sum_{j=1}^{N_{L-1}} g_{jk}^{(L)},
& \ell=L,
\end{cases}
\label{eq:ak_drn_piecewise}
\end{equation}
\begin{equation}
b_k^{(\ell)}=
\begin{cases}
-\displaystyle\sum_{j=1}^{N_{\ell-1}} g_{jk}^{(\ell)}\,v_j^{(\ell-1)}
\;-\;\displaystyle\sum_{m=1}^{N_{\ell+1}} g_{km}^{(\ell+1)}\,v_m^{(\ell+1)}
\;+\; I_k^{(\ell)},
& \ell=1,\dots,L-1,\\[10pt]
-\displaystyle\sum_{j=1}^{N_{L-1}} g_{jk}^{(L)}\,v_j^{(L-1)}
\;+\; I_k^{(L)},
& \ell=L.
\end{cases}
\label{eq:bk_drn_piecewise}
\end{equation}
\\
The exact coordinate update is characterized by the stationarity condition
\begin{equation}
\frac{\partial E_k^{(\ell)}}{\partial v_k^{(\ell)}}\!\bigl(v_k^{(\ell)}\bigr)
\;=\;
2a_k^{(\ell)}\,v_k^{(\ell)}
\;+\;
b_k^{(\ell)}
\;+\;
\frac{\partial \phi_k^{(\ell)}}{\partial v}\!\bigl(v_k^{(\ell)}\bigr)
\;=\; 0.
\label{eq:stationarity_Ek_drn}
\end{equation}
\\
For hidden layers $\ell=1,\dots,L-1$, we have
$\frac{\partial \phi_k^{(\ell)}}{\partial v}(v)= i_k^{(\ell)}(v)$, hence
\begin{equation}
2a_k^{(\ell)}\,v_k^{(\ell)}
\;+\;
b_k^{(\ell)}
\;+\;
i_k^{(\ell)}\!\bigl(v_k^{(\ell)}\bigr)
\;=\; 0.
\label{eq:1d_kcl_update_drn_hidden}
\end{equation}
\\
At the output layer $\ell=L$, there is no nonlinearity, so
\begin{equation}
2a_k^{(L)}\,v_k^{(L)} \;+\; b_k^{(L)} \;=\; 0.
\label{eq:1d_kcl_update_drn_output}
\end{equation}

\subsection*{Bidirectional amplification}

\begin{figure}
    \centering
    \includegraphics[width=0.5\linewidth]{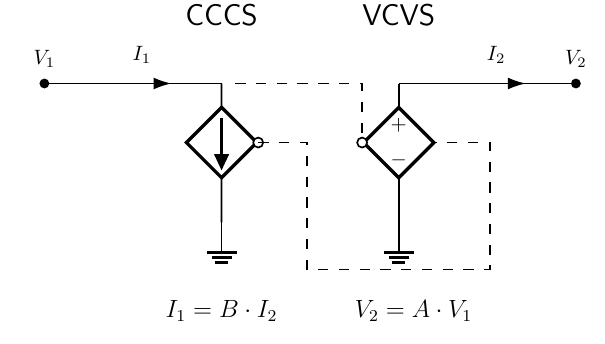}
    \caption{Schematic of a bidirectional amplifier. VCVS amplifies voltage in one direction, while CCCS amplifies current in the other direction.}
    \label{fig:bidiramp}
\end{figure}

In deep DRNs, voltage signals attenuate as they propagate through successive layers. To mitigate this signal attenuation while  preserving the network's bidirectionality, ref.~\cite{Kendall2020}
introduced a bidirectional amplifier between adjacent layers. The amplifier, schematized in Figure~\ref{fig:bidiramp}, consists of a
voltage-controlled voltage source (VCVS), which amplifies the forward voltage
by a factor \(A\), and a current-controlled current source (CCCS), which amplifies
the backward current by a factor \(B\). In ref.~\cite{Kendall2020}, the
reciprocal choice \(B=1/A\) was used. Here, we allow $A$ and $B$ to be chosen independently and
show how the resulting rescaling can be incorporated into the energy function of the network and the CD framework.

The energy function of the amplified DRN can be rewritten as
\begin{equation}
E(v)
\;=\;
\frac{1}{2}\sum_{\ell=1}^{L}\left(\frac{B}{A}\right)^{\ell-1}
\sum_{j=1}^{N_{\ell-1}}\sum_{k=1}^{N_{\ell}}
g^{(\ell)}_{jk}\,\bigl(\alpha_\ell\,v^{(\ell-1)}_{j}-v^{(\ell)}_{k}\bigr)^2
\;+\;
\sum_{\ell=1}^{L-1}\left(\frac{B}{A}\right)^{\ell-1}
\sum_{k=1}^{N_{\ell}}
\phi^{(\ell)}_{k}\!\bigl(v^{(\ell)}_{k}\bigr),
\label{eq:drn_energy_with_nl_vk_scaled_layerwise_input_not_amplified}
\end{equation}
with
\begin{equation}
\alpha_\ell=
\begin{cases}
1, & \ell=1,\\
A, & \ell\ge 2
\end{cases}.
\label{eq:alpha_def}
\end{equation}
where the nonlinear dissipation potential on each branch is
\begin{equation}
\phi^{(\ell)}_{k}(v)
\;:=\;
\int_{0}^{v} i^{(\ell)}_{k}(u)\,\mathrm{d}u.
\label{eq:phi_def_vk_scaled_drn}
\end{equation}
We then define
% For a node k in layer \ell, define "pre" = connections to next layer (\ell+1)
% and "post" = connections to previous layer (\ell-1).

\begin{align}
a_{\mathrm{pre},k}^{(\ell)}
&:= \sum_{m=1}^{N_{\ell+1}} g^{(\ell+1)}_{k m},
&
b_{\mathrm{pre},k}^{(\ell)}
&:= \sum_{m=1}^{N_{\ell+1}} g^{(\ell+1)}_{k m}\,v^{(\ell+1)}_{m},
\label{eq:apre_bpre}
\\[4pt]
a_{\mathrm{post},k}^{(\ell)}
&:= \sum_{j=1}^{N_{\ell-1}} g^{(\ell)}_{j k},
&
b_{\mathrm{post},k}^{(\ell)}
&:= \sum_{j=1}^{N_{\ell-1}} g^{(\ell)}_{j k}\,v^{(\ell-1)}_{j}.
\label{eq:apost_bpost}
\end{align}
and the stationarity condition is given by
\begin{equation}
\frac{\partial E}{\partial v_k^{(\ell)}}= \Bigl(a_{\mathrm{post},k}^{(\ell)}+\frac{B}{A}\,\alpha_{\ell+1}^{2}\,a_{\mathrm{pre},k}^{(\ell)}\Bigr)\,v_{k}^{(\ell)}
\;-\;
\Bigl(\alpha_{\ell}\,b_{\mathrm{post},k}^{(\ell)}+\frac{B}{A}\,\alpha_{\ell+1}\,b_{\mathrm{pre},k}^{(\ell)}\Bigr)
\;+\;
i_{\mathrm{nonlin}}^{(\ell)}\!\bigl(v_{k}^{(\ell)}\bigr)
\;=\;0.
\label{eq:cd_minimal_form}
\end{equation}

\section*{Data availability}
The Digits dataset is available through scikit-learn, and MNIST is available through standard machine-learning dataset repositories. The curated numerical data for the figures and tables is included in the repository at
\url{https://github.com/filip137/nonlinear-drn-paper}.

\section*{Code availability}
The source code needed to reproduce the results is available at\\
\url{https://github.com/filip137/nonlinear-drn-paper}.

\section*{Acknowledgements}

This work was supported by a France 2030 government grant managed by the French National Research Agency (ANR-22-PEEL-0010) and the European Research Council grant GRENADYN (101020684).

\section*{Author contributions}
F.O., J.G. and D.Q conceived the study. J.G. and D.Q. supervised the work. F.O. achieved the work. All authors regularly discussed the results. All authors contributed to write the manuscript.

\section*{Competing interests}
The authors declare no competing interests.

\FloatBarrier
\newpage
% Start supplementary numbering
\setcounter{section}{0}
\setcounter{figure}{0}
\setcounter{table}{0}
\setcounter{equation}{0}

\renewcommand{\thesection}{S\arabic{section}}
\renewcommand{\thefigure}{S\arabic{figure}}
\renewcommand{\thetable}{S\arabic{table}}
\renewcommand{\theequation}{S\arabic{equation}}
\section{Supplementary Material - Derivations}
\subsection{Lambert-$W$ derivations}
\subsubsection{Single Shockley diode}
\label{sec:supp_shockley_lambert}
For the single Shockley diode, the stationarity condition of the CD subproblem is
\begin{equation}
2a_k v_k + (b_k-I_s) + I_s e^{-v_{\mathrm{off}}/v_T}\exp\!\Big(\frac{v_k}{v_T}\Big)=0.
\label{eq:shockley_lambert_form}
\end{equation}
We define
\begin{equation}
A=2a_k,\qquad B=b_k-I_s,\qquad C=I_s e^{-v_{\mathrm{off}}/v_T},\qquad D=\frac{1}{v_T},
\label{eq:shockley_ABCD}
\end{equation}
so that \eqref{eq:shockley_lambert_form} becomes
\begin{equation}
A v_k + B + C e^{D v_k}=0
\qquad\Longleftrightarrow\qquad
A v_k + B = -C e^{D v_k}.
\label{eq:AvB_eq_exp}
\end{equation}
We now rewrite \eqref{eq:AvB_eq_exp} in the canonical form
\begin{equation}
z\,e^{z}=x
\qquad\Longleftrightarrow\qquad
z = W(x).
\label{eq:lambert_standard}
\end{equation}
We introduce
\begin{equation}
u := -D v_k - \frac{D B}{A}.
\label{eq:lambert_u_def_shockley}
\end{equation}
Then
\[
A v_k + B = -\frac{A}{D}\,u,
\qquad
e^{D v_k} = e^{-u}\,e^{-DB/A}.
\]
Substituting these identities into \eqref{eq:AvB_eq_exp} gives
\begin{equation}
u\,e^{u} = \frac{DC}{A}\,e^{-DB/A}.
\label{eq:lambert_canonical_shockley}
\end{equation}
Applying the principal branch \(W_0\), defined by \(W_0(x)e^{W_0(x)}=x\), yields
\[
u = W_0\!\left(\frac{DC}{A}\,e^{-DB/A}\right).
\]
Solving for \(v_k\) therefore gives
\begin{equation}
v_k^\star
=
-\frac{B}{A}
-\frac{1}{D}\,
W_0\!\left(
\frac{DC}{A}\,e^{-DB/A}
\right).
\label{eq:generic_solution_shockley}
\end{equation}
Finally, substituting \eqref{eq:shockley_ABCD} into \eqref{eq:generic_solution_shockley} gives
\begin{equation}
v_k^\star
=
-\frac{b_k-I_s}{2a_k}
-
v_T\,W_0\!\left(
\frac{I_s}{2a_k v_T}\,
\exp\!\Big(
-\frac{v_{\mathrm{off}}}{v_T}
-\frac{b_k-I_s}{2a_k v_T}
\Big)
\right).
\label{eq:supp_shockley_vstar_derivation}
\end{equation}
This reproduces Eq.~\eqref{eq:supp_shockley_vstar} in the main text.

\subsection{Piecewise Lambert-$W$ update for the double Shockley (antiparallel) diode}
\label{sec:supp_double_shockley_lambert}
For the double Shockley diode, and under the approximation that only a single diode is conducting at the time, the stationarity condition away from \(v_k=0\) is
\begin{equation}
\frac{\partial E}{\partial v_k}(v_k)
\;=\;
2a_k v_k+b_k+i_{\mathrm{LS}}(v_k)
\;=\;0.
\label{eq:supp_double_diode_stationarity_LS}
\end{equation}

On each branch, the second derivative is strictly positive:
\[
v_k>0:\quad
\frac{\partial^2 E}{\partial v_k^2}
=
2a_k+\frac{I_s}{v_T}\exp\!\Big(\frac{v_k-v_{\mathrm{off},1}}{v_T}\Big)>0,
\qquad
v_k<0:\quad
\frac{\partial^2 E}{\partial v_k^2}
=
2a_k+\frac{I_s}{v_T}\exp\!\Big(\frac{-(v_k+v_{\mathrm{off},2})}{v_T}\Big)>0.
\]
Hence \(E\) is strictly convex on each open branch. The one-sided derivatives at \(v_k=0\) are
\begin{equation}
\left.\frac{\partial E}{\partial v_k}\right|_{0^+}
=
b_k
+
I_s\exp\!\Big(-\frac{v_{\mathrm{off},1}}{v_T}\Big),
\qquad
\left.\frac{\partial E}{\partial v_k}\right|_{0^-}
=
b_k
-
I_s\exp\!\Big(-\frac{v_{\mathrm{off},2}}{v_T}\Big).
\label{eq:supp_dEdv_at_zero_sided}
\end{equation}
Thus, the derivative is strictly increasing across the two branches, with an upward jump at \(v_k=0\), and the global minimizer is unique. The signs of the one-sided derivatives determine its location:
\begin{equation}
\begin{aligned}
&\left.\frac{\partial E}{\partial v_k}\right|_{0^+}< 0
\;\Longrightarrow\;
v_k^\star>0,\\[2pt]
&\left.\frac{\partial E}{\partial v_k}\right|_{0^-}> 0
\;\Longrightarrow\;
v_k^\star<0,\\[2pt]
&\left.\frac{\partial E}{\partial v_k}\right|_{0^-}\le 0 \le
\left.\frac{\partial E}{\partial v_k}\right|_{0^+}
\;\Longrightarrow\;
v_k^\star=0.
\end{aligned}
\label{eq:branch_selection_simple}
\end{equation}
Using \eqref{eq:supp_dEdv_at_zero_sided}, this can be written equivalently as
\[
\begin{aligned}
&b_k < - I_s\exp\!\Big(-\frac{v_{\mathrm{off},1}}{v_T}\Big)
\;\Longrightarrow\; v_k^\star>0,\\
&b_k > \;\; I_s\exp\!\Big(-\frac{v_{\mathrm{off},2}}{v_T}\Big)
\;\Longrightarrow\; v_k^\star<0,\\
&- I_s\exp\!\Big(-\frac{v_{\mathrm{off},1}}{v_T}\Big)
\;\le\; b_k \;\le\;
I_s\exp\!\Big(-\frac{v_{\mathrm{off},2}}{v_T}\Big)
\;\Longrightarrow\; v_k^\star=0.
\end{aligned}
\]
 
Once the correct branch is known, \eqref{eq:supp_double_diode_stationarity_LS} reduces to a single-exponential equation of the form
\begin{equation}
A\,x+B+C\,e^{D x}=0,
\qquad A\neq 0,\; D\neq 0,
\label{eq:supp_lambert_generic_form}
\end{equation}
whose solution on the principal branch is
\begin{equation}
x^\star
=
-\frac{B}{A}
-\frac{1}{D}\,
W_0\!\left(
\frac{D C}{A}\,e^{-D B/A}
\right).
\label{eq:supp_lambert_generic_solution}
\end{equation}

\paragraph{Positive branch (\(v_k\ge 0\)).}
On the positive branch,
\[
2a_k v_k+b_k+I_s\exp\!\Big(\frac{v_k-v_{\mathrm{off},1}}{v_T}\Big)=0
\;\Longleftrightarrow\;
A\,v_k+B+C\,e^{D v_k}=0
\]
with
\[
A=2a_k,\qquad
B=b_k,\qquad
C=I_s e^{-v_{\mathrm{off},1}/v_T},\qquad
D=\frac{1}{v_T}.
\]
Substituting these quantities into \eqref{eq:supp_lambert_generic_solution} gives
\begin{equation}
v_k^\star
=
-\frac{b_k}{2a_k}
-\;v_T\,W_0\!\left(
\frac{I_s}{2a_k v_T}\,
\exp\!\Big(
-\frac{v_{\mathrm{off},1}}{v_T}
-\frac{b_k}{2a_k v_T}
\Big)
\right),
  \qquad v_k^\star>0.
\label{eq:supp_double_diode_LS_vstar_pos}
\end{equation}

\paragraph{Negative branch (\(v_k<0\)).}
On the negative branch,
\[
2a_k v_k+b_k-I_s\exp\!\Big(\frac{-(v_k+v_{\mathrm{off},2})}{v_T}\Big)=0
\;\Longleftrightarrow\;
A\,v_k+B+C\,e^{D v_k}=0
\]
with
\[
A=2a_k,\qquad
B=b_k,\qquad
C=-I_s e^{-v_{\mathrm{off},2}/v_T},\qquad
D=-\frac{1}{v_T}.
\]
Substituting these quantities into \eqref{eq:supp_lambert_generic_solution} gives
\begin{equation}
v_k^\star
=
-\frac{b_k}{2a_k}
+\;v_T\,W_0\!\left(
\frac{I_s}{2a_k v_T}\,
\exp\!\Big(
-\frac{v_{\mathrm{off},2}}{v_T}
+\frac{b_k}{2a_k v_T}
\Big)
\right),
\qquad v_k^\star<0.
\label{eq:supp_double_diode_LS_vstar_neg}
\end{equation}

\FloatBarrier
\section{Supplementary material - Solver Details}
\label{sec:supp_timing_analysis}
\subsection{Overrelaxed diode solver}

In our work we also consider the overrelaxed CD update. After computing the exact CD update \(v_{\mathrm{CD}}\), we set the next iterate to
\[
v^{\mathrm{new}} = v^{\mathrm{old}} + \omega \bigl(v_{\mathrm{CD}} - v^{\mathrm{old}}\bigr).
\]
Thus, \(\omega=1\) recovers plain CD, \(\omega>1\) gives overrelaxation, and \(\omega<1\) gives underrelaxation. In our work we use overrelaxation. We apply it for the single-diode and double-diode solvers. The minimizer also supports an optional reject/backtracking safeguard for \(\omega>1\), although we do not use it in practice in the runs reported here.
 
In the double-diode exponential sweep, overrelaxation substantially reduces the number of fixed sweeps needed to reach a given SPICE-relative error. The first sweep budget at which the node-weighted cross-layer relative \(L_1\) P90 error falls below \(10^{-4}\) is \(64\) for \(\omega=0.8\), \(32\) for \(\omega=1.0\), \(16\) for \(\omega=1.2\), and \(12\) for \(\omega=1.4\). For the \(2\times 10^{-5}\) threshold, the corresponding budgets are \(64\), \(32\), \(32\), and \(16\). Since one overrelaxed sweep has nearly the same cost as one standard sweep, this improves simulation speed.

\begin{figure}[t]
    \centering
    \includegraphics[width=0.82\linewidth]{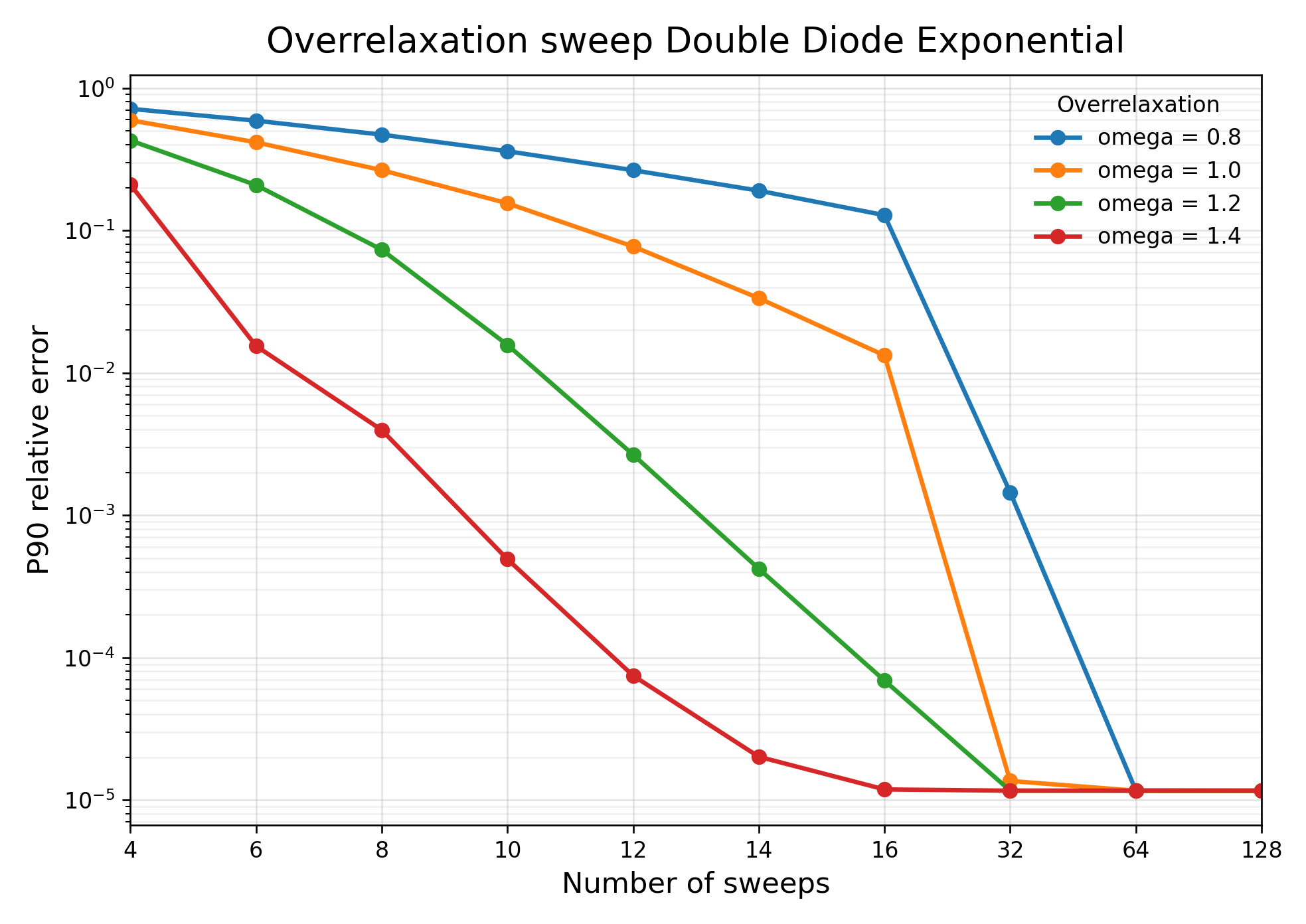}
    \caption{P90 node-weighted cross-layer relative \(L_1\) error versus CD sweeps for several overrelaxation factors in the three-hidden-layer, width-128 double-Shockley Digits network.}
    \label{fig:overrelaxation_iterations}
\end{figure}
\subsection{CD runtime decomposition}
\label{sec:supp_timing_runtime_decomposition}

The total CD runtime has two sources. The first is the number of odd-even sweeps required to satisfy the stopping criterion. This is a global convergence effect, and it is partly captured by the average-Hessian contraction proxy \(\rho(T)\) discussed in Supplementary Sec.~\ref{app:proxy}. We report these as \emph{outer sweeps}.

The second component is the work performed within each sweep. Every outer sweep contains many nonlinear coordinate updates. Although each such update has an analytical solution in terms of the Lambert \(W\) function, the implementation clips the Lambert argument for numerical stability. The clipped Lambert step is therefore only an initial approximation, which we then refine with a few Newton steps. We report the number of these as \emph{inner iterations}.

Both of these are reported for all runs in the Supplementary Table~\ref{tab:timing_runtime_decomposition}. 

\subsubsection{Clipped Lambert solve and Newton polish}
\label{sec:supp_clipped_lambert}

To interpret the inner-iteration counts, it is helpful to examine a single nonlinear coordinate solve. For the double-diode updater, each coordinate update reduces to one of the two scalar equations
\begin{equation}
2av+b+I_s\exp\!\left(\frac{v-v_{\mathrm{off}}}{V_T}\right)=0,
\label{eq:dd_forward_local}
\end{equation}
or
\begin{equation}
2av+b-I_s\exp\!\left(-\frac{v+v_{\mathrm{off}}}{V_T}\right)=0,
\label{eq:dd_reverse_local}
\end{equation}
depending on the active diode branch. All quantities are evaluated in \texttt{float64}.

In exact arithmetic, both equations can be solved in closed form using the Lambert \(W\) function. In practice, however, the Lambert argument has the form \(z=C\exp(\eta)\), and the exponent \(\eta\) can become numerically large. To avoid overflow and loss of stability, we clip this exponent to 165 before forming \(z\).

For the forward branch, we write
\begin{equation}
A_{+}=\frac{b}{2a},
\qquad
z_{+}=
\frac{I_s}{2aV_T}
\exp\!\left(
\operatorname{clip}\!\left(
-\frac{A_{+}+v_{\mathrm{off}}}{V_T}
\right)\right),
\qquad
v_{+}^{(0)}=-A_{+}-V_T W_0(z_{+}),
\label{eq:dd_forward_lambert}
\end{equation}
and for the reverse branch,
\begin{equation}
A_{-}=\frac{b}{2a},
\qquad
z_{-}=
\frac{I_s}{2aV_T}
\exp\!\left(
\operatorname{clip}\!\left(
\frac{A_{-}-v_{\mathrm{off}}}{V_T}
\right)\right),
\qquad
v_{-}^{(0)}=-A_{-}+V_T W_0(z_{-}).
\label{eq:dd_reverse_lambert}
\end{equation}

Because this clipping perturbs the exact closed-form solution, we apply a short Newton polish after the Lambert initialization. For the forward branch, we define
\begin{equation}
f_{+}(v)=2av+b+I_s\exp\!\left(\operatorname{clip}\!\left(\frac{v-v_{\mathrm{off}}}{V_T}\right)\right),
\label{eq:dd_forward_residual}
\end{equation}
and update
\begin{equation}
v \leftarrow v-\frac{f_{+}(v)}{f_{+}'(v)}.
\label{eq:dd_forward_newton}
\end{equation}
Similarly, for the reverse branch,
\begin{equation}
f_{-}(v)=2av+b-I_s\exp\!\left(\operatorname{clip}\!\left(-\frac{v+v_{\mathrm{off}}}{V_T}\right)\right),
\label{eq:dd_reverse_residual}
\end{equation}
with update
\begin{equation}
v \leftarrow v-\frac{f_{-}(v)}{f_{-}'(v)}.
\label{eq:dd_reverse_newton}
\end{equation}
The reported inner iterations are the Newton steps used.
\subsubsection{Odd--even sweep map and local convergence rate}
\label{app:proxy}
To understand the variations in the number of outer sweeps, it is convenient to view the odd--even CD solver as a nonlinear iteration map on the free voltages.
 
We can write the free node voltages, in the odd/even order, as
\[
v=(v_o,v_e).
\]
One odd--even sweep consists of two exact CD block minimizations,
\begin{align}
v_o^{k+1} &= \arg\min_{u_o} E(u_o,v_e^k), \label{eq:nonlinear_odd_update_app}\\
v_e^{k+1} &= \arg\min_{u_e} E(v_o^{k+1},u_e). \label{eq:nonlinear_even_update_app}
\end{align}
Equivalently, we can define
\[
\psi_o(v_e):=\arg\min_{u_o} E(u_o,v_e),
\qquad
\psi_e(v_o):=\arg\min_{u_e} E(v_o,u_e).
\]
Then one full sweep defines the nonlinear map
\begin{equation}
\Phi(v_o,v_e)
:=
\bigl(\psi_o(v_e),\ \psi_e(\psi_o(v_e))\bigr).
\label{eq:full_sweep_map_app}
\end{equation}
 
At a converged state \(v^\star=(v_o^\star,v_e^\star)\), we have
\[
v^\star=\Phi(v^\star).
\]
Writing the error as \(e^k:=v^k-v^\star\) and expanding around \(v^\star\) gives
\begin{equation}
e^{k+1}
=
D\Phi(v^\star)e^k+o(\|e^k\|).
\label{eq:local_jacobian_sweep_map_app}
\end{equation}
Near \(v^\star\), the error is governed to first order by \(D\Phi(v^\star)\). 
Thus the local asymptotic convergence rate is determined by the spectral radius
\[
\rho(D\Phi(v^\star))
:=
\max_{\lambda\in\sigma(D\Phi(v^\star))} |\lambda|,
\]
where \(\sigma(D\Phi(v^\star))\) denotes the set of eigenvalues of \(D\Phi(v^\star)\).
Thus, when \(\rho(D\Phi(v^\star))<1\), the number of sweeps needed to reduce the error to a fixed tolerance scales as
\[
k \propto \frac{1}{-\log \rho(D\Phi(v^\star))}.
\]
To make \(D\Phi(v^\star)\) explicit, we differentiate the functions that update blocks \(\psi_o\) and \(\psi_e\). These maps are defined implicitly by the stationarity conditions
\begin{equation}
\nabla_o E(\psi_o(v_e),v_e)=0,
\label{eq:odd_stationarity_app_text}
\end{equation}
and
\begin{equation}
\nabla_e E(v_o,\psi_e(v_o))=0.
\label{eq:even_stationarity_app_text}
\end{equation}
At the fixed point \(v^\star=(v_o^\star,v_e^\star)\), let
\[
H_{oo}:=\nabla_{oo}^2 E(v^\star),\qquad
H_{oe}:=\nabla_{oe}^2 E(v^\star),\qquad
H_{eo}:=\nabla_{eo}^2 E(v^\star),\qquad
H_{ee}:=\nabla_{ee}^2 E(v^\star),
\]
so that,
\[
H^\star=\nabla^2 E(v^\star)
=
\begin{pmatrix}
H_{oo} & H_{oe}\\
H_{eo} & H_{ee}
\end{pmatrix}.
\]
Differentiating \eqref{eq:odd_stationarity_app_text} with respect to \(v_e\) at \(v_e^\star\) gives
\[
H_{oo}\,D\psi_o(v_e^\star)+H_{oe}=0,
\]
and differentiating \eqref{eq:even_stationarity_app_text} with respect to \(v_o\) at \(v_o^\star\) gives
\[
H_{eo}+H_{ee}\,D\psi_e(v_o^\star)=0.
\]
Hence
\[
D\psi_o(v_e^\star)=-H_{oo}^{-1}H_{oe},
\qquad
D\psi_e(v_o^\star)=-H_{ee}^{-1}H_{eo}.
\]
Since
\[
\Phi(v_o,v_e)=\bigl(\psi_o(v_e),\ \psi_e(\psi_o(v_e))\bigr),
\]
the chain rule gives
\[
D\Phi(v^\star)
=
\begin{pmatrix}
0 & -H_{oo}^{-1}H_{oe}\\
0 & H_{ee}^{-1}H_{eo}H_{oo}^{-1}H_{oe}
\end{pmatrix}.
\]
The nonzero eigenvalues of \(D\Phi(v^\star)\) are exactly those of
\begin{equation}
T^\star
:=
H_{ee}^{-1}H_{eo}H_{oo}^{-1}H_{oe}.
\label{eq:Tstar_definition_app}
\end{equation}
\(T^\star\) is the reduced error-propagation matrix associated with one
odd--even sweep. When \(\rho(T^\star)<1\), the fixed point is locally linearly
stable, and smaller values of \(\rho(T^\star)\) correspond to faster asymptotic
decay of the error. For a fixed stopping tolerance, the expected number of sweeps
therefore scales with
\[
k \propto
\frac{1}{-\log \rho(T^\star)}.
\]
 
For the DRN used in our experiments, the energy function is given by
\begin{equation}
E(v)
=
\frac{1}{2}
\sum_{\ell=1}^{L} \left(\frac{B}{A}\right)^{\ell-1}
\sum_{j=1}^{N_{\ell-1}}\sum_{k=1}^{N_\ell}
g_{jk}^{(\ell)}
\bigl(\alpha_\ell v_j^{(\ell-1)}-v_k^{(\ell)}\bigr)^2
+
\sum_{\ell=1}^{L-1} \left(\frac{B}{A}\right)^{\ell-1} \sum_{k=1}^{N_\ell}
\phi_k^{(\ell)}\!\bigl(v_k^{(\ell)}\bigr)
+
\sum_{k=1}^{N_L} I_k^{(L)}\,v_k^{(L)},
\label{eq:drn_energy_proxy_app}
\end{equation}
where
\[
\alpha_\ell=
\begin{cases}
1, & \ell=1,\\
A, & \ell\ge 2.
\end{cases}
\]
 
The energy function has three terms; the quadratic term related to the linear conductive coupling between the layer, the non-linear term related to the non-linearities connecting the node voltage to the ground and the linear term connected to the current sources.
Since the current source term is linear, it does not contribute to the Hessian. The Hessian is thus given by
\begin{equation}
H(v)=\nabla^2 E(v)=H_{\mathrm{lin}}+D(v),
\label{eq:drn_hessian_structure_app}
\end{equation}
where \(H_{\mathrm{lin}}\) is the contribution of the linear conductances and \(D(v)\) is the contribution of the grounded nonlinearities.
 
Because each layer is coupled only to its immediate neighbors, \(H_{\mathrm{lin}}\) is block tridiagonal:
\begin{equation}
H_{\mathrm{lin}}
=
\begin{pmatrix}
M_1 & -\left(\frac{B}{A}\right)\alpha_2 G^{(2)} & & \\
-\left(\frac{B}{A}\right)\alpha_2 (G^{(2)})^\top & M_2 & \ddots & \\
& \ddots & \ddots & -\left(\frac{B}{A}\right)^{L-1}\alpha_L G^{(L)}\\
& & -\left(\frac{B}{A}\right)^{L-1}\alpha_L (G^{(L)})^\top & M_L
\end{pmatrix}.
\label{eq:drn_hlin_blocks_app}
\end{equation}
Here
\[
G^{(\ell)}=[g_{jk}^{(\ell)}]\in\mathbb{R}^{N_{\ell-1}\times N_\ell}
\]
denotes the weight matrix between layers \(\ell-1\) and \(\ell\). The diagonal entries of \(M_\ell\) collect the sum of linear conductances connected to each node
\begin{align}
(M_\ell)_{kk}
&=
\left(\frac{B}{A}\right)^{\ell-1}
\sum_{j=1}^{N_{\ell-1}} g_{jk}^{(\ell)}
+
\left(\frac{B}{A}\right)^\ell \alpha_{\ell+1}^2
\sum_{m=1}^{N_{\ell+1}} g_{km}^{(\ell+1)},
&& \ell=1,\dots,L-1,\\
(M_L)_{kk}
&=
\left(\frac{B}{A}\right)^{L-1}
\sum_{j=1}^{N_{L-1}} g_{jk}^{(L)}.
\end{align}

The grounded nonlinearities \(\phi_k^{(\ell)}(v_k^{(\ell)})\) depend only on a single node voltage, so their contribution to the Hessian is diagonal, with diagonal blocks given by the differential conductances:
\begin{equation}
D(v)
=
\operatorname{blkdiag}\!\Bigl(
\operatorname{Diag}\!\bigl(i^{(1)\prime}(v^{(1)})\bigr),
\frac{B}{A}\operatorname{Diag}\!\bigl(i^{(2)\prime}(v^{(2)})\bigr),
\dots,
\left(\frac{B}{A}\right)^{L-2}\operatorname{Diag}\!\bigl(i^{(L-1)\prime}(v^{(L-1)})\bigr),
0
\Bigr).
\label{eq:drn_nonlinear_hessian_app}
\end{equation}
For a given sample \(i\), the local asymptotic convergence of the odd--even solver is governed by the Hessian evaluated at that sample's converged state \(v_i^\star\). Since in our experiments we report the average number of outer sweeps over the validation set, we define a single average Hessian for each run by replacing the sample-dependent nonlinear differential conductances with their average over the converged validation states:
\begin{equation}
H_{\mathrm{avg}}
=
H_{\mathrm{lin}}
+
\operatorname{blkdiag}\!\Bigl(
c_1\operatorname{Diag}\!\bigl(\overline{i^{(1)\prime}(v^{(1)})}\bigr),
\dots,
c_{L-1}\operatorname{Diag}\!\bigl(\overline{i^{(L-1)\prime}(v^{(L-1)})}\bigr),
0
\Bigr),
\label{eq:average_hessian_app}
\end{equation}
where \(c_\ell=(B/A)^{\ell-1}\), and the overline denotes an elementwise average
over the converged validation states.

Using the same odd/even block convention as above, the corresponding
average-Hessian sweep matrix is
\begin{equation}
T
:=
\mathcal T(H_{\mathrm{avg}})
=
H_{ee}^{-1}H_{eo}H_{oo}^{-1}H_{oe}.
\label{eq:T_definition_app}
\end{equation}
This gives an average-Hessian estimate of the local contraction factor,
\(\rho(T)\). Since the number of sweeps scales as \(1/(-\log \rho)\) in the
linearized regime, we use
\begin{equation}
\frac{1}{-\log \rho(T)}
\label{eq:inv_log_rho_proxy_app}
\end{equation}
as the scalar predictor of the average number of outer sweeps.

\begin{figure}[t]
    \centering

    \begin{subfigure}[t]{0.82\linewidth}
        \centering
        \includegraphics[width=\linewidth]{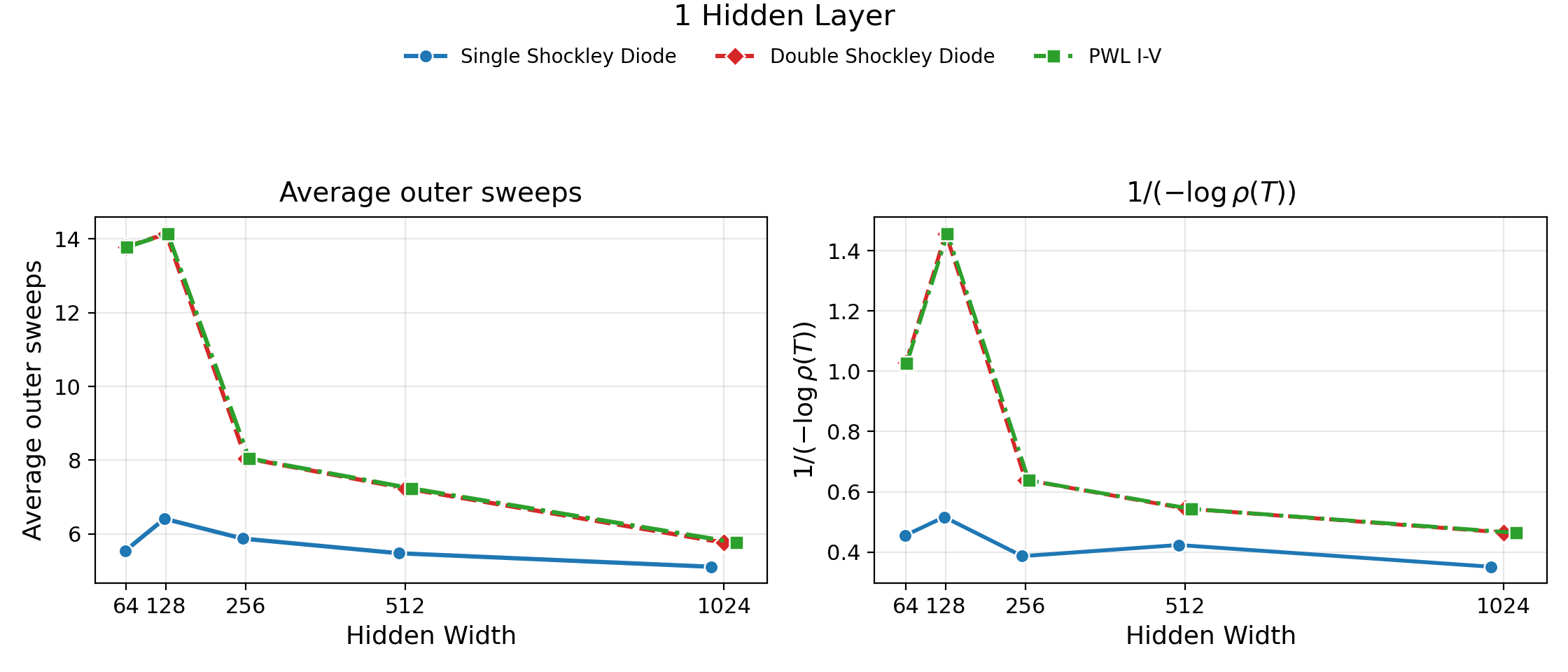}
        \caption{DRN with 1 Hidden Layer.}
        \label{fig:inv_log_rho_hidden1}
    \end{subfigure}

    \vspace{0.8em}

    \begin{subfigure}[t]{0.82\linewidth}
        \centering
        \includegraphics[width=\linewidth]{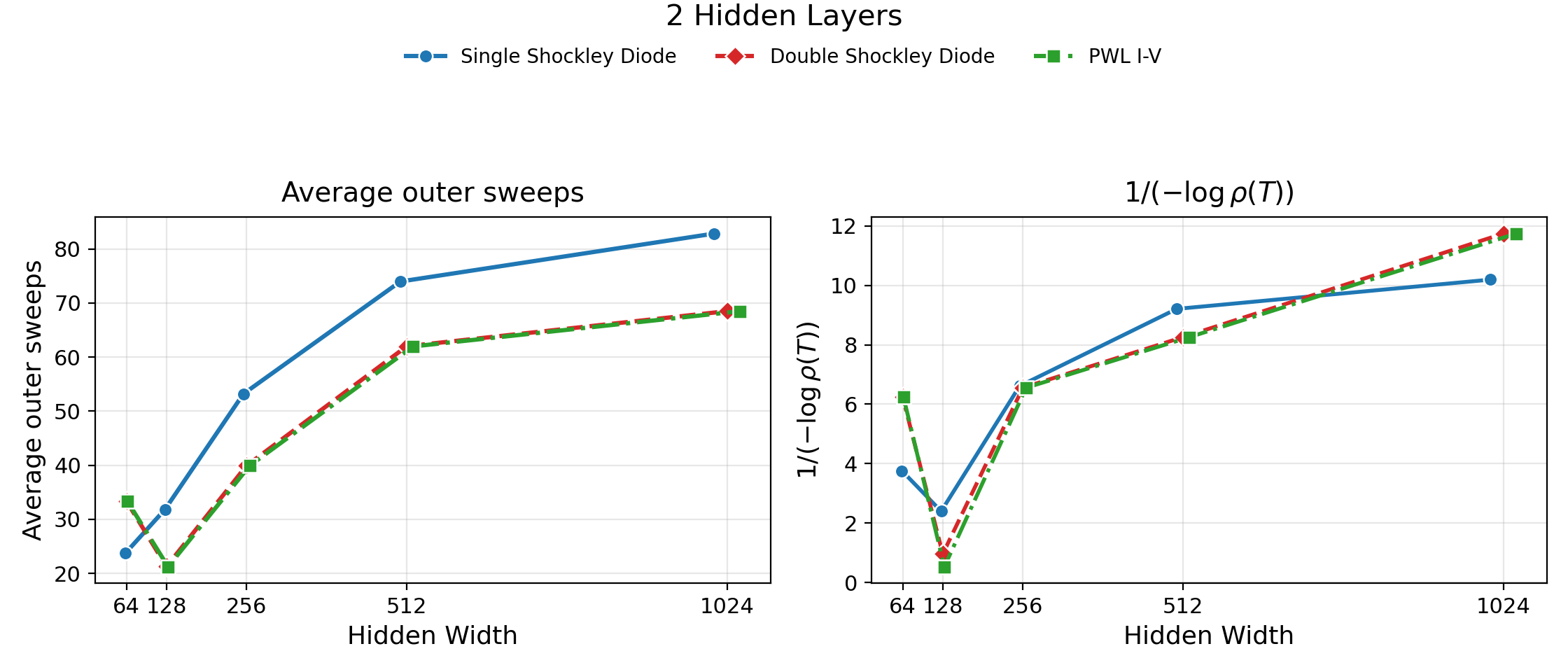}
        \caption{DRN with 2 Hidden Layers.}
        \label{fig:inv_log_rho_hidden2}
    \end{subfigure}

    \vspace{0.8em}

    \begin{subfigure}[t]{0.82\linewidth}
        \centering
        \includegraphics[width=\linewidth]{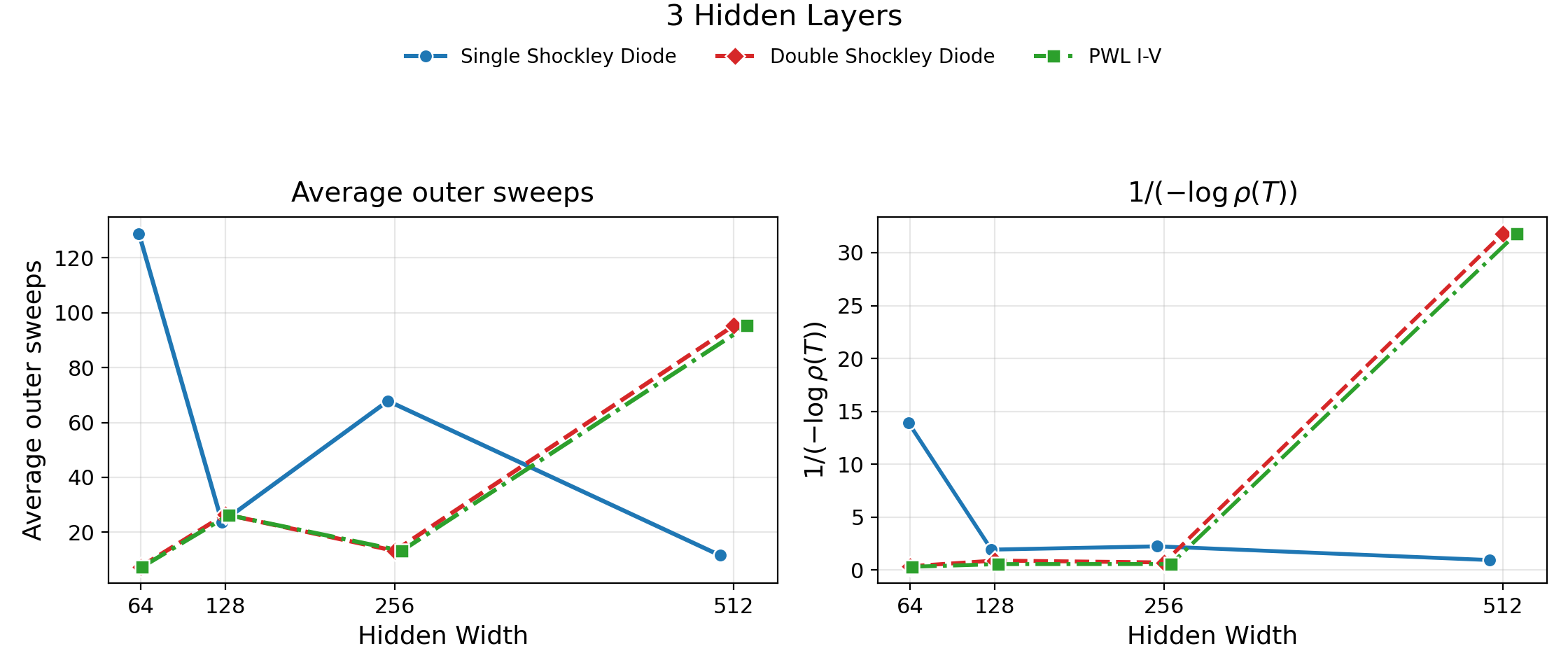}
        \caption{DRN with 3 Hidden Layers.}
        \label{fig:inv_log_rho_hidden3}
    \end{subfigure}

    \caption{Average outer sweeps and the contraction proxy \(1/(-\log \rho(T))\) for different layers widths. Each point is one Digits DRN configuration at the indicated hidden-layer depth, across nonlinearities and widths.}
    \label{fig:inv_log_rho_hidden123}
\end{figure}

In the Figures~\ref{fig:inv_log_rho_hidden1}, \ref{fig:inv_log_rho_hidden2} and \ref{fig:inv_log_rho_hidden3} this spectral radius is compared with the average number of outer sweeps for DRNs with one, two, and three hidden layers.

\iffalse
\begin{table}[H]
\centering
\caption{Pearson and Spearman correlations between \(1/(-\log \rho(T))\) and average outer sweeps, computed over the model-level points in Fig.~\ref{fig:inv_log_rho_hidden123}.}
\label{tab:model_level_corr}
\begin{tabular}{lcc}
\hline
Group & Pearson & Spearman \\
\hline
DRN with 1 Hidden Layer & 0.956 & 0.961 \\
DRN with 2 Hidden Layers & 0.897 & 0.963 \\
DRN with 3 Hidden Layers & 0.785 & 0.839 \\
\hline
\end{tabular}
\end{table}
\fi

\clearpage
\subsection{Timing summary tables}
\label{sec:supp_timing_tables}
Supplementary Tables~\ref{tab:timing_runtime_decomposition}
and~\ref{tab:timing_accuracy_summary}
provide the data from the timing experiments.
Supplementary Table~\ref{tab:timing_runtime_decomposition}
reports total runtimes together with outer-sweep and inner-iteration counts.
Supplementary Table~\ref{tab:timing_accuracy_summary}
reports the error between CD and SPICE.

\begin{landscape}
{\scriptsize
\setlength{\tabcolsep}{5pt}
\renewcommand{\arraystretch}{1.05}
\begin{longtable}{lccrrrrrr}
\caption{Runtime decomposition for the CPU timing runs included in the solver comparison. Outer sweeps denote the average number of equilibrium sweeps per sample. Inner sweeps denote the average number of local nonlinear-solver iterations per call. The runs for which SPICE netlist could not be generated are marked with dashes.}\label{tab:timing_runtime_decomposition}\\
\toprule
Nonlinearity & H & Width & CD [s] & SPICE [s] & Speedup & Sweeps avg. & Inner avg./call & Total inner\\
\midrule
\endfirsthead
\multicolumn{9}{l}{\tablename~\thetable\ (continued)}\\
\toprule
Nonlinearity & H & Width & CD [s] & SPICE [s] & Speedup & Sweeps avg. & Inner avg./call & Total inner\\
\midrule
\endhead
\midrule
\multicolumn{9}{r}{Continued on next page}\\
\endfoot
\bottomrule
\endlastfoot
Single Shockley & 1 & 64 & 7.246 & 59.305 & 8.2 & 11.972 & 1.181 & 10,176\\
Single Shockley & 1 & 128 & 7.679 & 30.891 & 4.0 & 9.000 & 2.000 & 12,960\\
Single Shockley & 1 & 256 & 5.889 & 69.386 & 11.8 & 5.878 & 2.000 & 8,464\\
Single Shockley & 1 & 512 & 6.013 & 198.342 & 33.0 & 5.469 & 1.990 & 7,836\\
Single Shockley & 1 & 1024 & 6.005 & 793.847 & 132.2 & 5.106 & 1.999 & 7,349\\
Single Shockley & 2 & 64 & 17.869 & 76.661 & 4.3 & 15.172 & 1.404 & 30,666\\
Single Shockley & 2 & 128 & 17.492 & 477.836 & 27.3 & 14.878 & 1.084 & 23,213\\
Single Shockley & 2 & 256 & 24.606 & 4293.154 & 174.5 & 21.456 & 1.000 & 30,896\\
Single Shockley & 2 & 512 & 39.363 & 49270.400 & 1251.7 & 31.775 & 1.000 & 45,756\\
Single Shockley & 2 & 1024 & 48.479 & -- & -- & 27.586 & 1.000 & 39,724\\
Single Shockley & 3 & 64 & 162.036 & 186.352 & 1.2 & 80.794 & 1.532 & 267,400\\
Single Shockley & 3 & 128 & 48.861 & 1737.107 & 35.6 & 17.853 & 2.408 & 92,876\\
Single Shockley & 3 & 256 & 132.086 & 12971.061 & 98.2 & 67.783 & 1.365 & 199,839\\
Single Shockley & 3 & 512 & 46.419 & -- & -- & 17.000 & 1.640 & 60,221\\
\addlinespace
Double Shockley & 1 & 64 & 5.923 & 22.065 & 3.7 & 9.000 & 1.895 & 6,141\\
Double Shockley & 1 & 128 & 4.767 & 73.608 & 15.4 & 8.972 & 1.000 & 3,230\\
Double Shockley & 1 & 256 & 5.946 & 100.499 & 16.9 & 9.000 & 1.966 & 6,371\\
Double Shockley & 1 & 512 & 7.815 & 309.581 & 39.6 & 9.000 & 1.994 & 6,462\\
Double Shockley & 1 & 1024 & 8.324 & 1107.349 & 133.0 & 9.000 & 2.000 & 6,480\\
Double Shockley & 2 & 64 & 14.678 & 68.514 & 4.7 & 16.750 & 1.000 & 12,060\\
Double Shockley & 2 & 128 & 14.702 & 549.956 & 37.4 & 15.847 & 1.213 & 13,837\\
Double Shockley & 2 & 256 & 16.778 & 4312.234 & 257.0 & 18.033 & 1.000 & 12,984\\
Double Shockley & 2 & 512 & 29.478 & 51768.450 & 1756.2 & 26.522 & 1.000 & 19,096\\
Double Shockley & 2 & 1024 & 55.053 & -- & -- & 31.731 & 1.000 & 22,846\\
Double Shockley & 3 & 64 & 27.569 & 138.930 & 5.0 & 9.578 & 1.879 & 30,427\\
Double Shockley & 3 & 128 & 21.922 & 1183.510 & 54.0 & 16.158 & 1.389 & 24,245\\
Double Shockley & 3 & 256 & 47.417 & 14361.823 & 302.9 & 16.644 & 7.844 & 141,011\\
Double Shockley & 3 & 512 & 78.675 & -- & -- & 51.936 & 1.000 & 56,091\\
\addlinespace
PWL i(v) & 1 & 64 & 11.807 & 22.023 & 1.9 & 13.778 & 20.352 & 100,948\\
PWL i(v) & 1 & 128 & 2.950 & 39.071 & 13.2 & 14.142 & 1.024 & 5,211\\
PWL i(v) & 1 & 256 & 8.974 & 94.790 & 10.6 & 8.050 & 20.875 & 60,497\\
PWL i(v) & 1 & 512 & 9.070 & 311.140 & 34.3 & 7.233 & 21.082 & 54,898\\
PWL i(v) & 1 & 1024 & 10.487 & 1094.952 & 104.4 & 5.761 & 24.476 & 50,764\\
PWL i(v) & 2 & 64 & 11.105 & 61.481 & 5.5 & 33.400 & 1.000 & 24,048\\
PWL i(v) & 2 & 128 & 12.756 & 543.731 & 42.6 & 21.236 & 3.956 & 60,485\\
PWL i(v) & 2 & 256 & 15.340 & 4301.242 & 280.4 & 39.936 & 1.000 & 28,754\\
PWL i(v) & 2 & 512 & 29.918 & 52009.107 & 1738.4 & 61.994 & 1.000 & 44,636\\
PWL i(v) & 2 & 1024 & 54.844 & -- & -- & 68.481 & 1.000 & 49,306\\
PWL i(v) & 3 & 64 & 15.679 & 165.079 & 10.5 & 7.422 & 15.818 & 126,800\\
PWL i(v) & 3 & 128 & 30.622 & 1131.481 & 37.0 & 26.231 & 6.905 & 195,624\\
PWL i(v) & 3 & 256 & 12.472 & 13723.796 & 1100.4 & 13.167 & 3.643 & 51,805\\
PWL i(v) & 3 & 512 & 67.944 & -- & -- & 95.214 & 1.000 & 102,831\\
\end{longtable}
}
\end{landscape}

\clearpage
{\small
\setlength{\tabcolsep}{5pt}
\renewcommand{\arraystretch}{1.05}
\begin{longtable}{lccrrrr}
\caption{Accuracy and error summary for the CPU timing runs included in the solver comparison. The runs for which SPICE netlist could not be generated are marked with dashes.}\label{tab:timing_accuracy_summary}\\
\toprule
Nonlinearity & H & Width & MAE & p50 rel. err. & p90 rel. err. & p99 rel. err.\\
\midrule
\endfirsthead
\multicolumn{7}{l}{\tablename~\thetable\ (continued)}\\
\toprule
Nonlinearity & H & Width & MAE & p50 rel. err. & p90 rel. err. & p99 rel. err.\\
\midrule
\endhead
\midrule
\multicolumn{7}{r}{Continued on next page}\\
\endfoot
\bottomrule
\endlastfoot
Single Shockley & 1 & 64 & 7.29e-06 & 1.27e-05 & 1.55e-05 & 1.81e-05\\
Single Shockley & 1 & 128 & 1.86e-05 & 1.51e-05 & 1.61e-05 & 1.71e-05\\
Single Shockley & 1 & 256 & 1.88e-05 & 1.55e-05 & 1.64e-05 & 1.70e-05\\
Single Shockley & 1 & 512 & 5.60e-06 & 1.10e-05 & 1.17e-05 & 1.23e-05\\
Single Shockley & 1 & 1024 & 5.78e-06 & 1.08e-05 & 1.14e-05 & 1.20e-05\\
Single Shockley & 2 & 64 & 4.11e-06 & 1.18e-05 & 1.39e-05 & 1.63e-05\\
Single Shockley & 2 & 128 & 4.59e-06 & 9.68e-06 & 1.08e-05 & 1.21e-05\\
Single Shockley & 2 & 256 & 1.93e-06 & 1.82e-05 & 2.06e-05 & 2.21e-05\\
Single Shockley & 2 & 512 & 1.90e-06 & 3.17e-05 & 3.81e-05 & 4.11e-05\\
Single Shockley & 2 & 1024 & -- & -- & -- & --\\
Single Shockley & 3 & 64 & 4.92e-05 & 6.62e-05 & 9.13e-05 & 1.08e-04\\
Single Shockley & 3 & 128 & 1.80e-05 & 1.53e-05 & 1.60e-05 & 1.65e-05\\
Single Shockley & 3 & 256 & 3.52e-05 & 2.05e-05 & 9.90e-05 & 2.13e-04\\
Single Shockley & 3 & 512 & -- & -- & -- & --\\
\addlinespace
Double Shockley & 1 & 64 & 9.72e-06 & 1.25e-05 & 1.46e-05 & 1.58e-05\\
Double Shockley & 1 & 128 & 1.87e-06 & 8.69e-06 & 9.95e-06 & 1.07e-05\\
Double Shockley & 1 & 256 & 6.37e-06 & 1.13e-05 & 1.26e-05 & 1.34e-05\\
Double Shockley & 1 & 512 & 6.84e-06 & 1.16e-05 & 1.25e-05 & 1.30e-05\\
Double Shockley & 1 & 1024 & 7.88e-06 & 1.21e-05 & 1.27e-05 & 1.32e-05\\
Double Shockley & 2 & 64 & 1.83e-06 & 1.34e-05 & 1.76e-05 & 2.07e-05\\
Double Shockley & 2 & 128 & 1.41e-05 & 1.17e-05 & 1.25e-05 & 1.30e-05\\
Double Shockley & 2 & 256 & 1.56e-06 & 1.65e-05 & 2.02e-05 & 2.35e-05\\
Double Shockley & 2 & 512 & 1.76e-06 & 2.30e-05 & 2.69e-05 & 2.95e-05\\
Double Shockley & 2 & 1024 & -- & -- & -- & --\\
Double Shockley & 3 & 64 & 2.20e-05 & 1.21e-05 & 1.28e-05 & 1.32e-05\\
Double Shockley & 3 & 128 & 1.09e-05 & 1.10e-05 & 1.19e-05 & 1.32e-05\\
Double Shockley & 3 & 256 & 1.03e-05 & 1.21e-05 & 1.27e-05 & 1.32e-05\\
Double Shockley & 3 & 512 & -- & -- & -- & --\\
\addlinespace
PWL i(v) & 1 & 64 & 1.06e-05 & 1.37e-05 & 1.58e-05 & 1.75e-05\\
PWL i(v) & 1 & 128 & 2.96e-06 & 1.36e-05 & 1.68e-05 & 1.99e-05\\
PWL i(v) & 1 & 256 & 6.53e-06 & 1.17e-05 & 1.28e-05 & 1.35e-05\\
PWL i(v) & 1 & 512 & 6.91e-06 & 1.17e-05 & 1.26e-05 & 1.33e-05\\
PWL i(v) & 1 & 1024 & 7.87e-06 & 1.21e-05 & 1.27e-05 & 1.33e-05\\
PWL i(v) & 2 & 64 & 9.11e-06 & 6.85e-05 & 7.97e-05 & 8.67e-05\\
PWL i(v) & 2 & 128 & 1.49e-05 & 1.21e-05 & 1.36e-05 & 1.68e-05\\
PWL i(v) & 2 & 256 & 6.15e-06 & 6.39e-05 & 7.41e-05 & 8.30e-05\\
PWL i(v) & 2 & 512 & 6.24e-06 & 8.07e-05 & 9.14e-05 & 1.03e-04\\
PWL i(v) & 2 & 1024 & -- & -- & -- & --\\
PWL i(v) & 3 & 64 & 2.13e-05 & 1.20e-05 & 1.27e-05 & 1.37e-05\\
PWL i(v) & 3 & 128 & 1.25e-05 & 1.22e-05 & 1.47e-05 & 2.07e-05\\
PWL i(v) & 3 & 256 & 1.04e-05 & 1.14e-05 & 1.19e-05 & 1.23e-05\\
PWL i(v) & 3 & 512 & -- & -- & -- & --\\
\end{longtable}
}

%\clearpage
%\input{tables/table_timing_run_accuracies_input_gain_longtable.tex}

%\clearpage
%\input{tables/table_accuracy_ladder_timing.tex}

\clearpage

\FloatBarrier
\clearpage

\newpage
\bibliographystyle{unsrtnat}
\bibliography{refs}
\end{document}